\documentclass[UTF8,a4paper]{article}
\usepackage{amsmath}
\usepackage{graphicx}
\usepackage{subfigure}
\usepackage{epstopdf}
\usepackage{geometry}
\usepackage{ulem}
\usepackage{indentfirst}
\usepackage{amsfonts,amssymb,dsfont}
\usepackage{setspace}
\usepackage{threeparttable}
\usepackage{amsmath,mathtools,amsthm}
\usepackage{array}
\usepackage{extarrows}
\usepackage{upgreek}

\makeatletter
\renewcommand*\env@matrix[1][\arraystretch]{%
  \edef\arraystretch{#1}%
  \hskip -\arraycolsep
  \let\@ifnextchar\new@ifnextchar
  \array{*\c@MaxMatrixCols c}}
\makeatother
\begin{document}


\title{\bf Fermi-liquid behaviors in three-dimensional polaron system with tunable dispersion}
\author{Chen-Huan Wu
\thanks{chenhuanwu1@gmail.com}
\\College of Physics and Electronic Engineering, Northwest Normal University, Lanzhou 730070, China}

\maketitle
\vspace{-30pt}
\begin{abstract}
\begin{large} 

We investigate the electronic properties and electron correlations 
in a three-dimensional (3D) polaron system with tunable dispersion parameter in Fermi-liquid picture.
The polaronic coupling is considered as a weakly attractive bare coupling.
Both the single and many-polaron, zero and finite-temperature cases are considered in this paper.
The physical quantities, including the 
self-energy, spectral function, optical conductivity (both the planar and 3D one), free energy,
and momentum distribution function,
are studied and are been verified 
exhibit Fermi-liquid features.
Both the analytic derivation and numerical calculation are done in this paper
to make our results reliable.\\
\\

\end{large}

\end{abstract}
\begin{large}

\section{Introduction}

Fermi-liquid theory is important in condensed-matter physics.
It is also interesting to explore,
for a three-dimensional (3D) polaron system, under what conditions the Fermi-liquid theory is valid.
In fact, the emergence of Fermi-liquid behavior, local Fermi-liquid behavior, or non-Fermi-liquid behavior
are related to the disorders, temperature, elastic/inelastic scattering channels, impurity density, and even the
particle-hole symmetry/asymmetry.
By studying the electronic properties of 3D polaron system in Fermi-liquid picture
with a momentum-independent coupling,
a lot of interesting physical results can be found,
including the effects of dispersion parameters, temperature, and UV/IR cutoff, etc.,
and many of them can not be found in both the non-Fermi-liquid state and even the
normal Fermi-liquid states, due to the existence of polaronic interaction.

In this paper, we are going to discuss both the single polaron and many-polaron cases
(with low-carrier density $n_{i}$),
that leads to great difference between our results with that obtained in the normal metals,
incoherence heavy Fermion systems (strongly correlated materials), 
and the high-temperature superconducting materials.
The polaron is treated as a long-lived quasiparticle in this paper 
with slow momenta and current relaxation.
The slow relaxation indeed exhibits Fermi-liquid feature
but this does not immediately related to the perturbative treatment,
since although the effective chemical potential $\mu=n_{i}g_{b}$ (first order self-energy),
which increases linearly with the bare coupling $g_{b}$,
indeed obeys the perturbative treatment,
the second order self-energy (to $g_{b}^{2}$)
does not gives such chemical potential,
and thus becomes nonperturbative.
The non-self-consistent $T$-matrix approximation 
(by summing up the ladder diagrams) is used during the calculation of second order self-energy.

In the presence of dynamically generated mass term $m$ (in instantaneous approximation),
the small value of $n_{i}/m$ may leads to a small Drude weight (and Drude conductivity)
and a lower Fermi temperature ($T_{F}\sim 1/m\sim \partial^{2}\varepsilon/\partial p^{2}$).
Different to the other systems,
the Drude conductivity studied in this paper is nonzero at zero temperature even for quasiparticle energy
$\varepsilon\gg \mu_{i}$.
This is due to the polaronic coupling-induced momentum transfer between the impurity 
and the majority particle.
During the calculation of optical conductivity and density-of-states,
we use the Lorentzian representation (instead of the Gaussian)
to take into account a finite small disorder (smearing).
We note that a smaller relaxation rate ($\eta=1/\tau={\rm Im}Sigma$),
will leads to a larger value of optical conductivity.
The imaginary part of self-energy leads to the damping of polaron.
At finite temperature with weak disorder,
the Laudau damping will doninate over the diffusion one.

For non-self-consistent scattering $T$-matrix,
where the impurity propagator is undressed by the self-energy,
has the poles (whose positions are varies with the value of momenta)
correspond to the resonance structure in the polaron spectrum (real part of the self-energy).
It is found that the nonresonance process will renormalizes the effective mass of quasiparticle
while the resonance process will leads to the so-called Planckian behavior
in the low-doped semiconductor or other strange metals 
(with carrier concentration $n\lesssim 10^{20}$cm$^{-3}$).
The resonance feature may be reduced by the decrease of exponent of dispersion of the
minority component (due to the increase of interaction strength), 
i.e., $\varepsilon_{p}\sim p^{\alpha}$ with $\alpha\ll 1$.
That is because the low-energy density-of-states (near the neutral point where the Fermi surface is nearby)
will lowered by the decreasing $\alpha$.

The rest of the paper is organized as follows.
In Sec.2 we discuss the self-energy, free-energy and specific heat at Fermi-liquid state.
In Sec.3 we study the optical conductivity.
In Sec.4, many-polaron effects are studied, like the electronic transport, WF law, 
momentum distribution function, and many-body spectral function.
In Sec.5, we discuss the relation between the ladder approximation 
used in this paper with the vertex correction.
In Sec.6 we conclude the main novel results obtained in this paper.

\section{Fermi-liquid state}

\subsection{Fermi-liquid behavior in self-energy and spectral function}

As the polaron formation\cite{1,2} does not contains the component which has a macroscopic amount,
like the phonons or photons (in a cavity),
the total momentum of the weakly-coupled propagating pair is conservative,
and thus the transport relaxation time of polaron has
$1/\tau\sim \Sigma^{(1)}=ng_{b}$ where $\Sigma^{(1)}$ is the
 first order contribution to self-energy.
$g_{b}$ is the bare coupling strength which equals
to the vacuum scattering matrix $T_{v}$.
For temperature higher than the critical one ($T_{c}=W^{2}/g_{b}$ with $W$ the bandwidth)
the non-Fermi-liquid feature,
which is hidden by the instability at low temperature,
will leads to incoherence between the electronic excitations.
Note that the finite-temperature also affects the resonance structure of the polaron spectrum\cite{Field B}.

Also, the strong coupling, like in a strongly correlated metals, will
largely reduce the critical temperature, and thus leads to the
coherence to incoherence crossover,
which also happen in the Dirac/Weyl semimetallic states\cite{Fujioka J}.
In this paper, we focus on the low-temperature non-Fermi-liquid region,
and study the polaron dynamics with arbitrary order of dispersion.

We write the isotropic dispersion of the impurity (with down-spin) and majority (with up-spin) particles 
as (before collision) $\varepsilon_{p\downarrow}=Ap^{\alpha}$ and $\varepsilon_{k\uparrow}=Bk^{\beta}$, respectively.
Then the bare Green's function are 
$G_{p\downarrow}^{-1}=\omega+i0-\varepsilon_{p\downarrow}$
and 
$G_{k\uparrow}^{-1}=\Omega+i0-\varepsilon_{k\uparrow}$, respectively.
When the pair propagator is not considered,
the self-energy at zero-temperature limit reads
\begin{equation} 
\begin{aligned}
\Sigma_{0}=&\int\frac{d\Omega}{2\pi}\int\frac{d^{3}q}{(2\pi)^{3}}
G_{0\uparrow}(p-q,\omega-\Omega)g_{b}.
\end{aligned}
\end{equation}
where we use the isotropic dispersion since the interaction is short-ranged.
Here the bare coupling term equivalent to the $T$-matrix 
in first-order of ladder expansion,
i.e., $ng_{b}=T$ where $n$ is the density of majority particle
that excited out of the Fermi surface (but not excited out of the medium,
since we consider only the vertices with the excitations created from or ahinilate to the
majority component, and 
 donot consider the propagators of ingong and outgoing particles which are
absorbed into and emitted from the Fermi bath through the scattering, respectively,
 although they may important
in a dilute BEC at both zero- and low-temperature which is the so-called Belyaev-type diagram),
and here the $T$-matrix is close to the vacuum one.
The mean field energy shift $ng_{b}$
(first order contribution to the self-energy)
here, as a impurity potential model, 
is also applicable for a randomly distributed uncorrelated impurities system,
as apply in Ref.\cite{Culcer D,ch}.
Note that the density $n$ (as well as the Fermi momentum $p_{F}$) here 
will affected by the temperature and interaction strength just like in the ultracold liquid or gas:
$n\propto T_{c}/T(T\ll T_{c})$, $n\propto 1/({\rm ln}g_{b})$\cite{Sears V F,Schick M,Field B}.
Base the above expression, for order of majority dispersion $\beta=1,2$,
the self-energy can be exactly solved,
but for $\beta\ge 3$, we
can only obtain the series expansion of the self-energy in terms of the momentum of majority particles.
Since we focus on the impurity momentum-dependence (locally critical theory),
the integral over frequency $\Omega$ is not done.

Base on the diagrammatic approach,
the second-order contribution to the polaron self-energy can be written as
\begin{equation} 
\begin{aligned}
\Sigma(p,\omega)=&\int\frac{d\Omega}{2\pi}\int\frac{d^{3}q}{(2\pi)^{3}}
G_{0\uparrow}(p-q,\omega-\Omega)T(q,\Omega)\\
\approx &T(p,\omega)\\
=&\int\frac{d\Omega}{2\pi}\int\frac{d^{3}q}{(2\pi)^{3}}\frac{1}{g_{b}^{-1}-\Pi(p-q,\omega-\Omega)}\\
=&\int\frac{d\Omega}{2\pi}\int\frac{d^{3}q}{(2\pi)^{3}}(g_{b}+g_{b}^{2}\Pi(p-q,\omega-\Omega)+
g_{b}^{3}\Pi^{2}(p-q,\omega-\Omega)
+\cdot\cdot\cdot).
\end{aligned}
\end{equation}
where $\Pi(p-q,\omega-\Omega)$ is the impurity-majority pair propagator.
The approximation in second line is valid
since the vertices (vertex correction-like;
as we will further discuss in following) contained in first line of above expression (in right-hand-side)
are suppressed in the contact interaction approximation
and a low density of the majority particles in the region where $g_{b}\neq  0$. 
Such short-range interaction (thus with small size of 
bound state) and low-density feature make the scattering potential can be savely replaced by a
contact potential.
While for a large density of majority component,
the single-particle bubble should be taken into account,
i.e., the majorarity propagator $G_{0\uparrow}(p-q,\omega-\Omega)$
should be keeped.

The corresponding self-energies, pair propagators, and spectral functions are shown in Figs.1-8
as functions of momentum $p$ or frequency $\omega$.
The spectral funtion reads 
\begin{equation} 
\begin{aligned}
A(p,\omega)=-2{\rm Im}G(p,\omega)=\frac{2|{\rm Im}\Sigma(p,\omega)|}
{[\omega-\varepsilon_{p}-{\rm Re}\Sigma(p,\omega)]^{2}+[{\rm Im}\Sigma(p,\omega)]^{2}}.
\end{aligned}
\end{equation}
From the self-energies shown in Fig.4
and the spectral function shown in Fig.5,
we see that the first-order self-energy $\Sigma_{0}(p,\omega)$ exhibit Fermi-liquid feature for
$\alpha=1$,
and non-fermi-liquid feature for $\alpha\neq 1$.
The non-Fermi-liquid features for $\alpha\neq 1$ include
the large value of ${\rm Im}\Sigma_{0}$ at $\omega=0$ and its
sharp initial rise (in consistent with the feature of inelastic scattering) and subsequent flattening behavior,
i.e., ${\rm Im}\Sigma_{0}\sim 1/\omega$ in large $\omega$-region.
The incoherent non-Fermi-liquid behavior can also be seen from the 
large peak width of structureless spectral functions as shown in Fig.5,
where the $\delta$-function peak is absent (especially for $\alpha=\beta\neq 1$)
and the spectral weight is much less than the Fermi-liquid one.
The non-Fermi-liquid features stated here can also be found in Hubbard model\cite{Liebsch A},
but are different from that exhibited in the ferromagnetic or anti-ferromagnetic quantum critical point
\cite{Wang J R}.
Note that both the Fermi surface and electron spectral function
can be probed experimentally by the
photoemission.

While the second-order self-energy $\Sigma(p,\omega)$ (Fig.6) exhibit Fermi-liquid feature
no matter how large the $\alpha$ and $\beta$ are,
and the imaginary part of self-energy (damping rate)
obeys a power law $|{\rm Im}\Sigma|\sim\omega^{a}\ (a>1)$ in small-$\omega$ region.
However,
we note that there exists finite peak width for $\alpha=\beta=3$ and $\alpha=\beta=4$ cases,
but the violation to the normal Fermi-liquid theory is weaker than the marginal one 
(in which case ${\rm Im}\Sigma$ linear in $\omega$ and quasiparticle residue logarithmic in $\omega$),
since the peaks are mostly symmetry.
Thus we treat them as the Fermi-liquid states here.
By comparing the spectral functions in Fig.7 and Fig.8,
we see that the spectral function as a function of momentum $p$ has a larger peak-width than 
that as a function of energy,
which implies the energy is a better quantum number here than the momentum,
and also consistent with the above statement that
the energy conservation is better retained than the momentum conservation during
the impurity-majority scattering.
Besides, although not shown,
we note that, our calculation reveals also that
with the increase of coupling $g_{b}$,
the spectral function peaks will slightly broadened although it is very not obvious.

By studying the analytical expression of the self-energies,
we also found that the imaginary part of self-energies in large-$\omega$ obey 
${\rm Im}\Sigma_{0}\sim{\rm ln}(1+\frac{1}{\omega})$,
${\rm Im}\Sigma    \sim{\rm ln}(1+\frac{1}{\omega^{2}})$
for $\beta=1$,
and 
${\rm Im}\Sigma_{0}\sim\frac{{\rm ln}(\frac{1}{\omega})}
{\sqrt{\omega}}$,
${\rm Im}\Sigma\sim\frac{{\rm ln}(\frac{1}{\omega})}
{\sqrt{\omega}}$ for $\beta>1$.

\subsection{Generated mass term}

Due to the interaction, the impurity (fermion) will dynamically obtain a finite mass (gap equation),
which is possible to leads to the charge-density wave (CDW) state.
The dynamically generated mass term in instantaneous approximation
reads
\begin{equation} 
\begin{aligned}
m(p,\omega)=\int\frac{d\Omega}{2\pi}\int\frac{d^{3}q}{(2\pi)^{3}}
\frac{m(q,\Omega)}{\varepsilon_{\uparrow}(q)}T(p+q,\omega+\Omega).
\end{aligned}
\end{equation}
For large order $\beta$, the low-energy DOS
can be treated as a constant $m$, 
then after ignore the frequency variable $\Omega$,
the dynamical $m(p)$ can be simply written as 
\begin{equation} 
\begin{aligned}
m(p)=\frac{-4m\pi q^{3}\Sigma(P)
{}_{2}F_{1}[1,\frac{3}{\beta},1+\frac{3}{\beta},\frac{Bq^{\beta}}{\mu_{\uparrow}}]}{3\mu_{\uparrow}}.
\end{aligned}
\end{equation}
Since here the Lorentz invariance is broken,
the dependence of mass term on momentum and frequency (of impurity) can be treated 
separately.
We found from this expression that the mass term
is proportional to the impurity self-energy and inversely proportional to the
eigenenergy of majority particle.
This is similar to mass enhancement due to the electron-phonon coupling
as described by the Eliashberg theory\cite{Allen P B}.
Since we apply the gapless dispersion for both the impurity and majority particles in the begining,
the mass enhancement here (due to the interaction effect) is also the final dispersion gap.
As presented in Fig.9,
note that only the real part (at positive $p$) of the gap equation contributes to the dispersion gas, thus
we can see that,
the mass $m(p)$ 
becomes positive only in a finite range of momentum (impurity becomes heavier),
and for small (close to zero) and large $p$, $m(p)\le 0$ (impurity becomes lighter).
It is for sure that, for $p\rightarrow \infty$,
the mass tends to zero,
which implies that the interaction effect as well as the polaron dynamic vanishes.

As the dispersion order is treated experimental turnable here,
we can then obtain a multiple-band system with different bandwidths
which exhibit marginal Fermi liquid feature in a finite temperature range,
where the transport relaxation time has
$1/\tau\sim {\rm Im}\Sigma\propto{\rm max}[\varepsilon,T]$
and the quasiparticle weight has $Z^{-1}\sim{\rm In}\frac{\Lambda}{\varepsilon}$.
The incoherent part of spectrum becomes dominate over the coherent part 
($\delta$-function) in this case.
Note that for impurity scattering,
when the random quenches of the impurities are taken into account,
the momentum is nomore conservative during the scattering event (due to the finite momentum relaxation),
and thus the system exhibit non-Fermi liquid behavior even at zero temperature ($T=0$)\cite{Buterakos D}.
Although the impurity we discuss in this paper is mobile,
the scattering can still be treated as elastic as long as the transferred energy is low enough,
e.g., much less than the temperature scale.
When the elastic scattering by quenched impurities is weak enough (low-energy event) and can be treated as a
perturbation,
the system can still behaves like a Fermi-liquid one,
e.g., obeys the Wiedemann-Franz law or broadened Drude peak in Ferimi-liquid metal\cite{Mahajan R}.


\subsection{Self-energy, free-energy and specific heat at finite temperature}

The real part of polaron self-energy changes the impurity dispersion,
and thus changes the free energy at finite temperature
(Legendre transform of energy),
which can be obtained by carrying the summation over the Matsubara frequency of the 
bare propagator.
Note that here the finite temperature considered is much lower than the critical value 
$T_{c}=W^{2}/g_{b}$,
where $W$ is the renormalized bandwidth of impurity.
The free energy reads
\begin{equation} 
\begin{aligned}
F(T)=&-2T\sum_{m}\int\frac{d^{3}p}{(2\pi)^{3}}
{\rm ln}(\sqrt{\omega_{m}^{2}+E(p)})\\
=&-4T\int\frac{d^{3}p}{(2\pi)^{3}}
{\rm ln}(1+e^{-E(p)/T}),
\end{aligned}
\end{equation}
where $E(p)=Ap^\alpha-\mu_{i}-{\rm Re}\Sigma_{0}(p)$.
For simplest case, where the self-energy is calculated by Eq.(1)
but for finite temperature,
the self-energy reads
\begin{equation} 
\begin{aligned}
\Sigma_{0}(p)=&\sum^{N}_{m'=-N}\int\frac{d^{3}q}{(2\pi)^{3}}
\frac{g_{b}}{\omega_{m'}+\omega_{n}+2i0-B(p-q)^{\beta}+\mu_{m}}.
\end{aligned}
\end{equation}
$\omega_{m'}=(2m'+1)\pi T/\hbar$ and $\omega_{n}$ are the fermionic Matsubara frequencies.
In above expression,
the summation over the index $m'$
can be done as following,
\begin{equation} 
\begin{aligned}
\sum_{m'=-N}^{N} &
\frac{1}{\omega_{m'}+\omega_{n}+2i0-B(p-q)^{\beta}+\mu_{m}}\\
&=\sum_{c=1}^{2N-1}
\frac{2(\mu_{m}-B(p-q)^{\beta}+\omega_{n}+2i0)}{(\mu_{m}-B(p-q)^{\beta}+\omega_{n}+2i0)^{2}-(c\pi T)^{2}}
+\frac{1}{\mu_{m}-B(p-q)^{\beta}+\omega_{n}+2i0+(2N+1)\pi T}.
\end{aligned}
\end{equation}
where $c$ is the odd integer.
The result of the summation in the first term of
right-hand-side is rather complex which contains four polygamma functions.
But for $N\rightarrow\infty$,
we obtain a simpler result for the first term (and the second term becomes zero)
\begin{equation} 
\begin{aligned}
\sum_{m'=-N}^{N}
\frac{1}{\omega_{m'}+i0-B(p-q)^{\beta}+\mu_{m}}=-\frac{{\rm tan}\frac{\mu_{m}-B(p-q)^{\beta}+i0}{2T}}{2T}.
\end{aligned}
\end{equation}
Thus the self-energy becomes calculable,
at least for $\alpha=\beta=1$,
\begin{equation} 
\begin{aligned}
\Sigma_{0}(p)=-&\int\frac{d^{3}q}{(2\pi)^{3}}
\frac{g_{b}{\rm tan}\frac{\mu_{m}-B(p-q)^{\beta}+i0}{2T}}{2T}\\
=&\frac{-1}{(2\pi)^{3}}
\frac{2 g_{b} \pi}{T} ((i q^{3})/3 - (
   2 q^{2} T {\rm ln}[1 + e^{(i (\mu_{m} - B p + B q +\omega_{n} + 2 i 0))/T}])/B\\
& + (
   4 i q T^2 {\rm Li}_{2}[-e^{((i (\mu_{m} - B p + B q +\omega_{n}+ 2 i 0))/T)}])/B^{2} \\
& - (
   4 T^{3} {\rm Li}_{3}[-e^{((i (\mu_{m} - B p + B q +\omega_{n}+ 2 i 0))/T)}])/B^{3})\bigg|_{q}.
\end{aligned}
\end{equation}
As shown in the last panel of Fig.10,
the above result for finite-temperature self-energy is only correct in small-$p$ and $T\neq 0$ 
region.
We also notice that the self-energy obtained here exhibits repeating resonance
structure over all the range of $p$, which is physically impossible,
this is because we take $m'\rightarrow\infty$ during the summation,
and the repeating resonance structure will no more exists when we set a proper cutoff
during the summation.

Since only the real part of self-energy affects the impurity dispersion,
the third line in above expression can be ignored.
Note that during the calculation of free energy here,
we carry the series expansion of the integrand to first order of $p$ to simplify the calculation,
then the range of integration over momentum $p$ need to be around $p=0$ 
(i.e., $\lambda$ need to be small and close to the chemical potential
while the length scale $L$ need to be large enough) which is necessary to obtain
the result of power-law changes in the temperature.
So in the premise of $\Lambda\gtrsim \mu_{i}\sim\mu_{m}$ and $\Lambda\gg e^{-L}$,
we obtain that $F(T)\sim T$ in the case of $\alpha=\beta=1$ for large $T$.
Note that this linear-in-temperature free-energy 
(corresponds to a constant entropy) 
will be accurate in the higher temperature regime,
but for low-temperature,
the free energy dispersion will shows a power law dependence with exponent larger than 2,
and thus leads to an increasing specific heat 
($C_{v}=-T\frac{\partial ^{2}F(T)}{\partial T^{2}}$) 
and entropy ($S=-\frac{\partial F(T)}{\partial T}$) with temperature.
Here we note that, 
the specific heat usually behaves as $C\sim T^{x}$ with $x>3$\cite{Stewart G R}
in Fermi-liquid system where the relevant energy
scale is Fermi energy but not temperature ($T\ll \mu_{i}$),
and $C\sim T^{x}$ with $x<2$ in a broad temperature range for non-Fermi-liquid system,
including the topological quantum critical mode\cite{Wang J R2,Han S E},
while in the non-Fermi-liquid systems with the interacting phonons-induced linear-in-$T$ resistivity,
the specific heat may even with a exponent $0<x<1$ in the low-$T$ region and $x<0$ in higher $T$ region,
which is the well known $C\sim c_{1}-T{\rm ln}T$ behavior, 
as experimentally studied\cite{Rost A W,Mahajan R}.
Here $c_{1}$ is added to make sure $C$ vanishes at zero temperature limit,
since the entropy $S$ vanishes at such limit for any interacting system.

As shown in Fig.10,
since we have subtract the zero-temperature part of free energy,
the free energy here is zero at $T=0$,
and we can also see that the small $L$ will leads to wrong result.
While for large $L$ (at least larger than 5 here),
the change of $L$ does not affect the free energy anymore.
In the regime of temperature close to zero,
the resonance structure is originated from the self-energy resonance in unstable region.
For $\beta>1$ case, larger length scale $L$ (long wavelength) is needed for the calculation.
Through redious calculation (carry the series expansion until the order of $T^{4}$),
the free energy for $\beta=2$
can be obtained as $F(T)\sim c_{1}+(16\pi T+O(T^{5}))c_{2}$,
where $c_{1}$ is a small constant, and $c_{2}$ is a parameter slightly depend on temperature
(can approximately treated as a constant).
Thus for $\beta=2$, at less in a moderate temperature range,
the linear-in-temperature feature should still dominate.
Note that, for interacting case,
the first-order self-energy applied here (not the mean-field shift) 
as a function of temperature is shown in Fig.11(a)-(b),
which increases with increasing temperature
at low-$T$ regime and 
acts as $\Sigma_{0}(T)\sim {\rm ln}(1/T)$ at large temperature (especially for larger $\Lambda$ 
as shown in Fig.11(b)).
Here we found that at low temperature, the real part of self-energy is proportional to the
thermal energy as ${\rm Re}\Sigma\sim T$,
while the imaginary part of self-energy (see Fig.11(c)-(d))
has ${\rm Im}\Sigma\sim {\rm ln}T+c_{1}$
which is in contrast with the non-Fermi-liquid behavior
${\rm Im}\Sigma\sim T$.

While in noninteracting case for $\alpha=1$,
the free energy is obtained as
\begin{equation} 
\begin{aligned}
F(T)=&-4T\int\frac{d^{3}p}{(2\pi)^{3}}
{\rm ln}(1+e^{-\frac{Ap^{\alpha}-\mu_{i}}{T}})\\
=&
-16 \pi T ((
   p^{3} (A p + 4 T {\rm ln}[1 + e^{(\mu_{i} - A p)/T}] - 
      4 T {\rm ln}[1 + e^{(-\mu_{i} + A p)/T}]))/(12 T) \\
&- (
   p^{2} T {\rm Li}_{2}[-e^{((-\mu_{i} + A p)/T)}])/A + (
   2 p T^{2} {\rm Li}_{3}[ -e^{((-\mu_{i} + A p)/T)}])/A^{2} \\
&- (
   2 T^{3} {\rm Li}_{4}[ -e^{((-\mu_{i} + A p)/T)}])/A^{3}).
\end{aligned}
\end{equation}
For $\alpha=2$,
after use the series expansion
\begin{equation} 
\begin{aligned}
{\rm ln}(1+e^{-Ap^{2}/T})
={\rm ln}2-\frac{p^{2}}{2T}+\frac{p^{4}}{8T^{2}}+O(p^{6}),
\end{aligned}
\end{equation}
we obtain
\begin{equation} 
\begin{aligned}
F(T)\approx &
-16 \pi T \left[-(\frac{e^{(\mu_{i}/T)} p^{5} (-5 p^{2} + 14 (1 + e^{(\mu_{i}/T)}) T)}{
    70 (1 + e^{(\mu_{i}/T)})^{2} T^{2}}) + \frac{1}{3} p^{3} {\rm ln}[1 + e^{(\mu_{i}/T)}]\right]\bigg|_{p}.
\end{aligned}
\end{equation}
In comparasion to the interacting case with the polaron self-energy correction,
the free energy obtained in noninteracting case is strictly linear in
temperature as shown in third panel of Fig.10.
We found that the free energy is linear-in-temperature whatever the $\alpha$ is,
and the free energies obtained under different $\alpha$ is almost the same.

As we state above, the free-energy as a function of temperature shows linear-in-$T$ feature 
in high temperature regime and parabolic in low temperature regime,
and we consider only the first order contribution to self-energy here.
To see if this related to the temperature-dependence of self-energy within the polaron energy term,
we next apply the mean-field shift (mean-field polaron energy) $\Sigma_{0}=ng_{b}T$,
which is temperature independent.
Through calculation we obtain the free-energy as 
\begin{equation} 
\begin{aligned}
F(T)=&
16 \pi T ((
   p^{3} (p + 4 T {\rm ln}[1 + e^{-(p/T)}] - 4 T {\rm ln}[1 + e^{p/T}]))/(12 T) - \\
&   p^{2} T {\rm Li}_{2}[ -e^{p/T}] + 2 p T^{2} {\rm Li}_{3}[ -e^{p/T}] - 
   2 T^{3} {\rm Li}_{4}[ -e^{p/T}])\bigg|_{p},
\end{aligned}
\end{equation}
which is shown in Fig.12.
The free-energy here implies that,
at least the temperature-dependence of fisrt-order self-energy
is not the only factor that leads to the
features of free-energy stated above.

\section{Optical conductivity}

In a many-polaron system,
the optical conductivity can be observed in the presence of linear response
to a homogeneous electric field applied in a single direction (longitudinal) or
a monochromatically oscillating vector potential (transverse).
Firstly, the longitudinal conductivity (parallel to the $p_{x}-p_{y}$ plane
) reads
\begin{equation} 
\begin{aligned}
\sigma_{ii}(\omega)=\frac{1}{V\omega}\int^{\infty}_{0}dte^{i\omega t}\langle J_{i}(t)J_{j}(0)\rangle,
\end{aligned}
\end{equation}
where the bracket denote the averaging in grand canonical ensemble.
Since the current density operator at Fermi surface has $J=ev_{F}=e\frac{p_{F}}{m^{*}}$,
where the velocity operatore reads $v=i[H,r]/\hbar$,
the above expression can be rewritten as
\begin{equation} 
\begin{aligned}
\sigma_{ii}(\omega)=\frac{e^{2}}{V(m^{*})^{2}\omega^{3}}\int^{\infty}_{0}dte^{i\omega t}\langle F_{i}(t)F_{j}(0)\rangle.
\end{aligned}
\end{equation}
$F=-\dot{p}=i[H_{int},p]/\hbar$ is the external force operator,
where $H_{int}$ denotes the interacting part of the Hamiltonian:
\begin{equation} 
\begin{aligned}
H=\sqrt{N}\int\frac{d^{2}k}{(2\pi)^{2}}(n^{*}g_{p}c_{q}^{\dag}+ng_{p}^{*}c_{q}),
\end{aligned}
\end{equation}
with $n=\sum_{{\bf r}}e^{i{\bf p}\cdot{\bf r}}$ is the Fourier-transformed impurity density and $N$
is the density of majority particle.

Base on the Kubo formula,
the in-planar optical conductivity for a monochromatically oscillating vector potential reads
\begin{equation} 
\begin{aligned}
\sigma_{ij}(\omega)=ie^{2}\int\frac{d^{2}p}{(2\pi)^{2}}
\frac{(N_{F}(\varepsilon_{sp})-N_{F}(\varepsilon_{s'p'}))\langle s|v_{i}|s'\rangle\langle s'|v_{j}|s\rangle}
{(\varepsilon_{sp}-\varepsilon_{s'p'})(\omega+i0+\varepsilon_{sp}-\varepsilon_{s'p'})},
\end{aligned}
\end{equation}
where $s=\pm 1$ is the band indices, and $p'$ is close to $p$ due to the optical limit.
Note that here we consider an external light field or eletric field-induced 
polaronic coupling with extremely small but finite transferred momentum $q$,
thus it is indeed different from the optical conductivity simply single-particle excitation
(with zero transferred momentum in optical limit).
The effect of driving field here can still be seen in the following 
text when we calculate the effect of polaron-polaron interaction.
Through this expression, we know that,
in the cases of intraband transition or transition between two very close bands,
the optical conductivity does not related to the coupling constant and
chemical potentials at zero temperature
in which case the distribution functions in numerator
becomes the step function.
Here $v$ can be obtained through the Hamiltonian $H$,
where we treat the interacting part momentum-independently,
\begin{equation} 
\begin{aligned}
H=&H_{0}+H_{int}\\
\approx &
\begin{pmatrix}
0 & A(p_{x}-ip_{y})^{\alpha}\\
A(p_{x}+ip_{y})^{\alpha} & 0
\end{pmatrix}
+
ng_{b}.
\end{aligned}
\end{equation}
Note that when consider the particle-hole symmetry during the interband transition,
we have $\varepsilon_{sp}=-\varepsilon_{s'p}$ and $\varepsilon_{s(p-q)}=-\varepsilon_{s'p'}$,
while for the intraband transition or the
transition between two bands very close to each other,
we have $\varepsilon_{s(p-q)}=\varepsilon_{s'p'}\approx \varepsilon_{s'p}$.
We will calculate both these two cases in the following.
For isotropic dispersion,
we have $p_{x}=p_{y}=\frac{\sqrt{2}p}{2}$ where $p$ denote the planar momentum here.
Then the velocity matrix elements read (for interband transition)
\begin{equation} 
\begin{aligned}
\langle s|v_{x}|s'\rangle=&\frac{1}{p}((1 + i) (-1)^{3/4} 2^{-\frac{1}{2} - \frac{\alpha}{2}} A \alpha ((1 - i) p)^{
 \alpha/2} ((1 + i) p)^{\alpha/2})\delta_{s,s'-1},\\
\langle s'|v_{y}|s\rangle=&\frac{1}{p}((1 - i) (-1)^{1/4} 2^{-\frac{1}{2} - \frac{\alpha}{2}} A \alpha ((1 - i) p)^{
 \alpha/2} ((1 + i) p)^{\alpha/2})\delta_{s',s+1}.
\end{aligned}
\end{equation}
Note that the velocity matrices here are the $2\times 2$ diagonal matrices,
and we only use the positive element to calculate the conductivity.
The relation $\sum_{s}\langle s|v_{ij}|s\rangle
={\rm Tr}v_{ij}=v_{11}+v_{22},\ \sum_{ss'}\langle s|v_{ij}|s'\rangle
=\sum_{ss'}\langle s^{*}|v_{ij}|s'\rangle=ss'(v_{11}+v_{22})$
is used.
Although external field-induced energy difference $|Ap^{\alpha}-A(p-q)^{\alpha}|$ 
is taken into account,
we still use the approximated velocity matrix here:
$v_{i}=\langle s|\frac{\partial H_{s}}{\partial p}+\frac{\partial (H_{s}-H_{s'})}{\partial p}|s'\rangle
\approx\langle s|\frac{\partial H_{s}}{\partial p}|s'\rangle$.
By setting the transferred momentum $q$ small enough,
we obtain the optical conductivities as shown in Fig.13 (real part) and Fig.14 (imaginary part).

For three-dimensional case,
when donot consider the spin-orbit coupling (SOC),
the Hamiltonian reads
\begin{equation} 
\begin{aligned}
H=&
\begin{pmatrix}
ng_{b}+Cp_{z} & A(p_{x}-ip_{y})^{\alpha}\\
A(p_{x}+ip_{y})^{\alpha} & ng_{b}+Cp_{z}
\end{pmatrix}.
\end{aligned}
\end{equation}
and the corresponding velocity matrix elements read 
\begin{equation} 
\begin{aligned}
\langle s|v_{x}|s'\rangle=&\frac{1}{p}((1 + i) (-1)^{3/4} 2^{-(1/2) - \alpha/2} A \alpha ((1 - i) p)^{
 \alpha/2} ((1 + i) p)^{\alpha/2})\delta_{s,s'-1},\\
\langle s'|v_{y}|s\rangle=&\frac{1}{p}((1 - i) (-1)^{1/4} 2^{-(1/2) - \alpha/2} A \alpha ((1 - i) p)^{
 \alpha/2} ((1 + i) p)^{\alpha/2})\delta_{s',s+1}.
\end{aligned}
\end{equation}
Note that if the SOC is taken into accout,
the Hamiltonian becomes $H\otimes\begin{pmatrix}
S_{z} & S_{x}-iS_{y}\\
S_{x}+iS_{y} & S_{z}
\end{pmatrix}$, where $S_{i}$ is the components of spin/pseudospin operator\cite{Ezawa M},
or simply through the orthogonal transformation matrix\cite{Orlita M}.

As we donot consider the SOC here, the three-dimensional conductivity reads
\begin{equation} 
\begin{aligned}
\sigma(\omega)=-ie^{2}\int\frac{d^{2}p}{(2\pi)^{2}}\int\frac{dp_{z}}{2\pi}
\frac{(N_{F}(\varepsilon_{p})-N_{F}(\varepsilon_{p'}))\langle s|v_{i}s'\rangle\langle s'|v_{j}|s\rangle}
{(\varepsilon_{p}-\varepsilon_{p'})(\omega+i0+\varepsilon_{p}-\varepsilon_{p'})},
\end{aligned}
\end{equation}
where the eigenenergy here reads
\begin{equation} 
\begin{aligned}
\varepsilon_{p}=Ap^{\alpha}-\mu_{i}+ng_{b}+Cp_{z},\\
\varepsilon_{p'}=A(p-q)^{\alpha}-\mu_{i}+ng_{b}+C(p_{z}-q).
\end{aligned}
\end{equation}
As can be seen from the expression of planar and 3D optical conductivity, only
the interband optical conductivity is related to the chemical potential and $g_{b}$,
while the intraband optical conductivity is not.
This is also why the Pauli blocking is only possible during the interband transition,
i.e., a characteristic onset at $\omega=2\mu_{i}$.
The analytic result of 3D conductivity can be exactly obtained until $\alpha>3$
as shown in panels in the right-hand-side of Fig.13.
Similar to the 2D (planar) optical conductivity,
the 3D optical conductivity here has $\sigma_{xx}=\sigma_{yy}$.

Note that as we consider the gapless case here, if the 
the Pauli blocking effect is taken into account (for interband absorption),
the step function $\theta(\omega-2|\mu_{i}|)$ can be added to the expression of interband optical conductivity,
in the presence of weak disorder and within the first-order perturbation theory.
For smaller transferred momentum, the optical conductivity $\sigma_{xy}$ will becomes
higher.
From the first four panels of Fig.13, we see that for $s\neq s'$ (interband), 
the Drude peak is absent even in the presence of finite chemical potential and broadening 
(set as $\eta=1/\tau=0.01$
throughout this work),
and when integer factor $\alpha>2$,
there is a kink structure at finite frequency in the optical conductivity $\sigma_{xy}$
corresponding to the frequency required by interband transition,
while for $\alpha=1,2$,
the optical conductivity dispersion
acts like the case of $s=s'$ (decrease logarithmically)
and the kink structure is absent for nonzero frequency.
Also, we notice that the interband optical conductivities increase with the increasing 
$\alpha$ or $|g_{b}|$ (although not shown here) in the region $\mu_{i} \ll \omega\ll \omega_{\Lambda}=A\Lambda^{\alpha}$,
and with the increase of $\alpha$ or coupling,
the real part of optical conductivity converges more slowly.

As shown in the last four panels of Fig.13 (intraband),
the Drude peaks (broadened by the weak disorder) 
can be found from both the intraband planar and 3D $\sigma_{xy}$ close to zero frequency,
and they should be replaced by the delta-like peak in clean limit.
In non-Fermi-liquid systems, like the good metals with strong interactions\cite{Craco L,Mahajan R},
in the absence of quasiparticle (and thus the total momentum is possible to be conserved),
a sharp Drude peak can be found in the intraband optical conductivity.
This is obviously in contrast with the case we study here.
For $s=s'$,
both the planar and 3D conductivities increase with increasing $\alpha$
at large $\omega$ region.
An important difference from the results obtained by Fermi's golden rule in Ref.\cite{Devreese J T}
is the coupling-dependence of the real part of optical conductivity ${\rm Re}\sigma\propto V_{k}^{2}$,
this is essentially because we use the approximation of constant coupling in short-range limit,
that also leads to a Drude weight which is independent of the coupling.
Once the coupling parameter is related to the impurity momentum,
we always have $\sigma_{xy}\propto g_{p}^{2}$.
Note that since the spin-momentum locking is not being considered here,
the transverse conductivity has $\sigma_{xy}=\sigma_{yx}$
which is in contrast with the spin-charge correlated system\cite{Keser A C},
thus for the thermal average we have the expectation value of
conductivity $\langle {\bf \sigma}\rangle
={\rm Tr}[\hat{n}{\bf \sigma}]=2\sigma_{xy}$,
where the density operator $\hat{n}$ here is to diagonalize the conductivity.

We mainly focus on the physical characters in Fermi-liquid regime in this paper.
As can be seen from the optical conductivities obtained here,
they are very different from the ones obtained in non-Fermi-liquid regime,
which satisfy the relation $\sigma\sim \omega/T$,
i.e., linear with frequency of driving field at a certain temperature,
as we presented in detail in Ref.\cite{higher}.

Except the kink structrue and $g_{b}$-dependence,
the difference between the interband and intraband optical conductivities also includes 
the spectral weight obtained by optical sum rule.
For interband optical conductivity,
to fulfill the identity (obey the $f$-sum rule)
$\int^{\infty}_{0}\frac{d\omega}{\pi}\omega{\rm Re}\sigma(\omega)=\frac{e^{2}n}{2m^{*}}$ in Fermi-liquid theory,
the sum need to be performed over all bands while for the intraband optical conductivity
the sum need to be performed over only one band (i.e., the Drude weight).
The effective mass here is different from the one in bare Drude weight 
(in absence of self-energy and vertex 
correction and with the preserved Galilean invariance)\cite{Abedinpour S H,Maiti S}.
Note that the $f$-sum rule is applicable in RPA 
which stipulate the conservation of particle number and energy,
thus requires the momentum transfer fast enough (adiabatically).
That is satisfied in the optical limit,
where we have $\sum_{p}G(p-q)\approx \sum_{p}G(p)$ even in the unbounded region.
As can be seen through the above analysis,
the optical conductivities increase with increasing $\alpha$
especially at large $\omega$ region.
Although we use the one-polaron limit during the calculation,
we can predict that the optical conductivity is proportional to $\alpha$,
since higher $\alpha$ will increases the density-of-states ($\rho(\omega)$)
at low-energy region,
and thus the carrier density is also increased ($n=\int^{\mu}_{0}d\omega\rho(\omega)$),
which leads to higher optical conductivity in a many-polaron system
($\sigma(\omega)\propto ne^{2}/\omega$\cite{Tempere J}).
Here the density-of-states reads (with first-order self-energy)
\begin{equation} 
\begin{aligned}
\rho(\omega)=\int\frac{d^{3}p}{(2\pi)^{3}}
\delta(\omega-(Ap^{\alpha}-\mu_{i}+ng_{b})),
\end{aligned}
\end{equation}
and (with second-order self-energy)
\begin{equation} 
\begin{aligned}
\rho(\omega)=\int\frac{d^{3}p}{(2\pi)^{3}}
\delta(\omega-(Ap^{\alpha}-\mu_{i}+\Sigma(p,\omega))),
\end{aligned}
\end{equation}
which becomes depends on both the $\alpha$ and $\beta$.
Restricted to the case of first order self-energy,
we use the Lorentzians representation to deal with the delta function within the integral,
\begin{equation} 
\begin{aligned}
\delta(\omega-(Ap^{\alpha}-\mu_{i}+ng_{b}))=\frac{\eta}{\pi}
\frac{1}{(\omega-(Ap^{\alpha}-\mu_{i}+ng_{b}))^{2}+\eta^{2}},
\end{aligned}
\end{equation}
then we obtain
\begin{small}
\begin{equation} 
\begin{aligned}
\rho(\omega)=&
-\frac{1}{(2\pi)^{3}}
(\frac{2}{3}) i p^{3} [\frac{
   1 - (\frac{A p^{\alpha}}{-\mu_{i} + g_{b} n + A p^{\alpha} - \omega + i z})^{-\frac{3}{\alpha}}
      {}_{2}F_{1}[-(\frac{3}{\alpha}), -(\frac{3}{\alpha}), \frac{-3 + \alpha}{\alpha}, \frac{
      \mu_{i} - g_{b} n + \omega - i z}{\mu_{i} - g_{b} n - A p^{\alpha} + \omega - i z}]}{
   \mu_{i} - g_{b} n + \omega - i z} \\
&- \frac{
   1 - (-(\frac{A p^{\alpha}}{\mu_{i} - g_{b} n - A p^{\alpha} + \omega + i z}))^{-\frac{3}{\alpha}}
      {}_{2}F_{1}[-(\frac{3}{\alpha}), -(\frac{3}{\alpha}), \frac{-3 + \alpha}{\alpha}, \frac{
      \mu_{i} - g_{b} n + \omega + i z}{\mu_{i} - g_{b} n - A p^{\alpha} + \omega + i z}]}{
   \mu_{i} - g_{b} n + \omega + i z}]
\bigg|_{p=0}^{\Lambda}.
\end{aligned}
\end{equation}
\end{small}
where ${}_{2}F_{1}[x]$ is the hypergeometric function.
We show in Fig.16 the density-of-states as a function of $\omega$.
We found that for small $\omega$ the 
density-of-states (to first-order of self-energy)
decreases with the increase of order $\alpha$.
While for large $\omega$, we
see that the DOS is proportional to the $\alpha$,
which is consisitent our above statement:
the carrier concentration as well as the minimal conductivity is increased with
the increasing dispersion order $\alpha$.
But note that this is only valid in large $\omega$ region,
which is to make sure that the $\omega>\varepsilon_{p}$.
There is a collapse (at critical frequency $\omega_{c}$) in the DOS no matter what $\alpha$ is,
and the value of this critical frequency is increased with the increasing $\alpha$,
we can only make sure that $\rho(\omega>\omega_{c},\alpha=a+1)>\rho(\omega>\omega_{c},\alpha=a)$.
Note that here we consider the energy distribution of
single band density-of-states,
while for a multi-subband system (with the same $\alpha$ or not),
the weight of wave function of each subband need to be taken into account, as done in 
Ref.\cite{Woods B D}.

Additionaly,
we here provides another way to compute the imaginary part of conductivity which is
more effectively when deal with the discrete Matsubara sum at finite temperature.
When the many-body effect is not taken into account,
the real part of interband optical conductivity formula requires a delta-function,
\begin{equation} 
\begin{aligned}
{\rm Re}\sigma_{xy}(\omega)=&
-\pi e^{2}\int\frac{d^{2}p}{(2\pi)^{2}}
\frac{N_{F}(s\varepsilon_{sp})-N_{F}(s'\varepsilon_{s'p'})}{\varepsilon_{sp}-\varepsilon_{s'p'}}
\langle s|v_{i}|s'\rangle \langle s'|v_{j}|s\rangle\delta(\omega+(\varepsilon_{sp}-\varepsilon_{s'p'}))\\
=&-\pi e^{2}\frac{1}{(2\pi)^{2}}
\frac{
(A \Lambda^{2} \pi^{2} \delta[A q + w])
}{q},
\end{aligned}
\end{equation}
which can be obtained through the Sokhotski Plemelj theorem $1/(x\pm i\eta)=P(1/x)\mp i\pi\delta(x)$
where $P$ is the integral principal value.
Note that as the integrand is an even function,
the momentum $p$ is integrated over the range of zero to UV cutoff $\Lambda$.
With the increase of carrier density, the delta-function here in above expression should be replaced by the dynamical 
structure factor\cite{Tempere J},
which contains both the continuum part (damped)
and undamped part.
While for the intraband ${\rm Re}\sigma_{xx}$ and ${\rm Re}\sigma_{yy}$, we found they 
exactly in the same expression with ${\rm Re}\sigma_{xy}(\omega)$ calculated above,
that can also be verified through the result obtained in above as shown in Fig.13,
the only difference here is the Sokhotski Plemelj theorem used here requires the clean limit,
i.e., $\eta\rightarrow 0$, while we set a finite lifetime here.

Then the imaginary part of conductivity can be obtained through the Kramers-Kronig relations\cite{Lucarini V}
\begin{equation} 
\begin{aligned}
{\rm Im}\sigma(\omega)=&\frac{-1}{\pi}P\int^{\infty}_{-\infty}
d\omega'\frac{{\rm Re}\sigma(\omega')}{\omega'-\omega}\\
=&\frac{-2\omega}{\pi}P\int^{\infty}_{0}
d\omega'\frac{{\rm Re}\sigma(\omega')}{\omega^{'2}-\omega^{2}},
\end{aligned}
\end{equation}
Since when the integral here is convergent it converges to its principal value,
the intraband optical conductivity for $\alpha=1$ reads
\begin{equation} 
\begin{aligned}
{\rm Im}\sigma_{xy}(\Omega)
=&\frac{-2\Omega}{\pi}P\int^{\infty}_{0}
        d\omega\frac{{\rm Re}\sigma(\omega)}{\omega^{2}-\Omega^{2}}\\
=&\frac{-2\Omega}{\pi}\lim_{a\rightarrow\infty}\int^{a}_{0}
        d\omega\frac{{\rm Re}\sigma(\omega)}{\omega^{2}-\Omega^{2}}\\
=&\frac{-2\Omega}{\pi}\lim_{a\rightarrow\infty}
 (-\pi e^{2})\frac{1}{(2\pi)^{2}}
\frac{A \Lambda^{2} \pi^{2} (-1 + 2 \theta[a]) \theta[-A q - 
   a \theta[-a]] \theta[
  A q + a \theta[a]]}{q (A^{2} q^{2} - \Omega^{2})}\\
=&\frac{-2\Omega}{\pi}
 (-\pi e^{2})\frac{1}{(2\pi)^{2}}
(A \Lambda^{2} \pi^{2} \theta[-A q])/(A^{2} q^{3} - q \Omega^{2}),
\end{aligned}
\end{equation}
which turns out to be inverse proportional to the transferred momentum $q$,
and the dependence on $\Omega$ decreased with decreasing $q$.
For $\alpha>2$, the above procedure can be repeated.
Similarly, if we do not directly use the replacement $i\omega\rightarrow\omega+i0$
in the Green's function (as done in above section),
the real part of self-energy can still be obtain by 
firstly using the Lehmann spectral representation and then the Kramers-Kronig relation.
Through this expression,
we see that the imaginary part of optical conductivity vanishes in static limit (dc-conductivity)
unlike the real part,
which is also in consistent with the result of Ref.\cite{Vasko F T}.

For intraband optical conductivity,
the finite temperature effect can be taken into account by 
the relation of 
$\sigma(T)=\int d\omega\sigma(T=0)\frac{-\partial N_{F}(\omega)}{\partial \omega}$,
where the factor $\frac{-\partial N_{F}(\omega)}{\partial \omega}$
becomes $\delta(\mu-\omega_{p})$ at zero temperature limit.
This expression sometimes also used in the independent-particle approximation to calculate the interband
optical conductivity\cite{Shao Y}.

For the interband optical conductivity,
the finite temperature effect can be taken into account by 
consider the thermal factor 
\begin{equation} 
\begin{aligned}
[N_{F}(s\varepsilon_{p})-N_{F}(s'\varepsilon_{p'})]\delta(\omega-2\varepsilon_{p})=
\frac{{\rm sinh}\frac{\omega}{2T}}{{\rm cosh}\frac{\omega}{2T}+{\rm cosh}\frac{\mu_{i}}{T}}.
\end{aligned}
\end{equation}
If the small $q$ is taken into account,
the above equation can be rewritten as (to $O(q/T)$)
\begin{equation} 
\begin{aligned}
&[N_{F}(s\varepsilon_{p})-N_{F}(s'\varepsilon_{p'})]\delta(\omega+s\varepsilon_{p}-s'\varepsilon_{p-q})\\
=&
[N_{F}(s\varepsilon_{p})-N_{F}(s'\varepsilon_{p})]+\frac{A\alpha e^{s'\varepsilon_{p}}p^{\alpha-1}q}
{N_{F}^{2}(s'\varepsilon_{p})T}\\
=&
\frac{{\rm sinh}\frac{\omega}{2T}}{{\rm cosh}\frac{\omega}{2T}+{\rm cosh}[\frac{\mu_{i}}{T}
+\frac{A\alpha p^{\alpha-1}q\omega}{(ng_{b}+Ap^{\alpha})4T}]}\\
=&\frac{{\rm sinh}[\frac{w}{2 T}]}{
 {\rm cosh}[\mu_{i}/T] + 
  {\rm cosh}[\frac{w}{2 T}]} - \frac{(A \alpha p^{-1 + \alpha}
     w {\rm sinh}[\mu_{i}/T] {\rm sinh}[\frac{w}{2 T}]) q}{
 4 ((g_{b} n + A p^{\alpha}) T ({\rm cosh}[\mu_{i}/T] + {\rm cosh}[\frac{w}{2 T}])^2)}+O(q^{2}/T^{2}).
\end{aligned}
\end{equation}
To see the temperature effect,
we plot the interband optical conductivity $\sigma_{xy}$ in Fig.15.
We see that the optical conductivity decreases with the increasing temperature,
and for large enough temperature, it becomes a constant at nonzero frequency.
The resonance structure at low temperature can be seen at the frequency close to $2\mu_{i}$,
which becomes less pronounced at long-wavelength limit (vanishing $q$).
From Fig.15,
we can clearly see that, due to the isotropic dispersion,
we have $\sigma_{xy}=\sigma_{xx}=\sigma_{yy}$.
Due to the constant coupling, our result of finite-temperature optical conductivity
is, compared to the optical conductivity in the presence of
long-range Coulomb potential, more close to the optical conductivity in 
the presence of short-range impurity\cite{Abedinpour S H,Park S},
and have the relation 
${\rm Re}\sigma\sim ({\rm ln}\Lambda/\omega)^{-1}$
in the collisionless regime
$\Lambda\gg \omega\gg k_{B}T(\gg q)$ 
where the optical conductivity tends to a constant
(see also Ref.\cite{Sodemann I,Abedinpour S H}).
Note that in collisionless regime,
as the $\omega$-dependence of optical conductivity is largely lowered,
the Boltzmann equation can still has
$\frac{\partial N_{F}}{\partial p}=v\frac{\partial N_{F}}{\partial \varepsilon}\neq 0$
as the distribution function here contains the nonequilibrium term.

\section{Many-polaron effect}

\subsection{transport}

Consider the polaron-polaron interaction (with momentum transfer $q'$) into account,
the many-body Hamiltonian reads
\begin{equation} 
\begin{aligned}
H=\sum_{p}\varepsilon_{p}c^{\dag}_{p}c_{p}+\sum_{q}\varepsilon_{q}d^{\dag}_{q}d_{q}
+\frac{1}{2}\sum_{p,p',q'}g_{p-p}
 c^{\dag}_{p+q'}c^{\dag}_{p-q'}c_{p}c_{p'}
+\sum_{k,p,q}g_{b}
 d^{\dag}_{k+q}c_{p-q}^{\dag}c_{p}d_{k},
\end{aligned}
\end{equation}
where $g_{p-p}$ denotes the contact interaction between down-spin impurities (or polarons),
and $q'$ is the transferred momentum during the polaron-polaron interaction.
which can be obtained by carrying the unitary transformation 
to the low-energy effective Hamiltonian (as shown in above).
In this case, 
since the current (or current density) operator reads $\hat{J}_{ij}=ev\hat{n}(e<0)$
with the impurity (carrier) density operator (Fourier transformed)
here as $\hat{n}=\sum_{r}e^{i{\bf q}'\cdot{\bf r}}=\sum_{p}\psi^{\dag}_{p-q'}\psi_{p}$,
where $\psi_{p-q'}$ is the field operator 
(or the corresponding eigenvector) which has $\psi_{p-q'}=c_{p-q'}$ when we donot consider
the pseudospin or other flavors.
Then we have
\begin{equation} 
\begin{aligned}
\hat{J}_{ij}=&e\sum_{p}\psi^{\dag}_{p-q',i}v_{ij}\psi_{p,j}\\
=&e\sum_{p}c^{\dag}_{p-q',i}v_{ij}c_{p,j},
\end{aligned}
\end{equation}
which equals the charge $e$ times the velocity-density operator.
In following we omit the spatial components of velocity operator $i,j$ within the field operator $c$.
The continuity equation thus reads
\begin{equation} 
\begin{aligned}
i\partial_{t}\hat{n}=&[\hat{n},H_{0}]+[\hat{n},H_{int}]\\
=&q'\otimes
\begin{pmatrix}
0 & v_{ij}^{*}\sum_{r}e^{-i{\bf q}'\cdot{\bf r}}\\
 v_{ij}\sum_{r}e^{i{\bf q}'\cdot{\bf r}} & 0    
\end{pmatrix}
+
\sum_{k,p,q}g_{q}[\psi^{\dag}_{p-q'}\psi_{p},d^{\dag}_{k+q}c^{\dag}_{p-q}c_{p}d_{k}]
+
\sum_{p,p'}g_{q'}[\psi^{\dag}_{p-q'}\psi_{p},c^{\dag}_{p+q'}c^{\dag}_{p-q'}c_{p}c_{p'}]\\
=&q'\otimes
\begin{pmatrix}
0 & v_{ij}^{*}\sum_{r}e^{-i{\bf q}'\cdot{\bf r}}\\
 v_{ij}\sum_{r}e^{i{\bf q}'\cdot{\bf r}} & 0    
\end{pmatrix}
+
\sum_{k,p,q}g_{q}[c^{\dag}_{p-q'}c_{p},d^{\dag}_{k+q}c^{\dag}_{p-q}c_{p}d_{k}]
+
\sum_{p,p'}g_{q'}[c^{\dag}_{p-q'}c_{p},c^{\dag}_{p'+q'}c^{\dag}_{p-q'}c_{p}c_{p'}]\\
=&
\begin{pmatrix}
0 & q'v_{ij}^{*}\sum_{r}e^{-i{\bf q}'\cdot{\bf r}}\\
 q'v_{ij}\sum_{r}e^{i{\bf q}'\cdot{\bf r}} & 0    
\end{pmatrix}
+
\sum_{k,p,q}g_{q}d^{\dag}_{k+q}c^{\dag}_{p-q-q'}c_{p}d_{k}.
\end{aligned}
\end{equation}
where $H=H_{0}+H_{int}$ and since $n$ is no more a conserved quantity
when consider the interaction of $g_{b}=g_{q}$ and $g_{p-p}=g_{q'}$,
we have $[n,H]
\neq 0$.
The above expression can be obtained through the relation
\begin{equation} 
\begin{aligned}
[c^{\dag}_{e}c_{f},g_{abcd}c^{\dag}_{a}c^{\dag}_{b}c_{c}c_{d}]
=\frac{1}{2}[g_{abec}c^{\dag}_{a}c^{\dag}_{b}c_{c}c_{f}
            -g_{abce}c^{\dag}_{a}c^{\dag}_{b}c_{c}c_{f}]
+\frac{1}{2}[g_{bcfa}c^{\dag}_{e}c^{\dag}_{a}c_{b}c_{c}
            -g_{bcaf}c^{\dag}_{e}c^{\dag}_{a}c_{b}c_{c}],
\end{aligned}
\end{equation}
as
\begin{equation} 
\begin{aligned}
[c^{\dag}_{p-q'}c_{p},d^{\dag}_{k+q}c^{\dag}_{p-q}c_{p}d_{k}]
=&d^{\dag}_{k+q}c^{\dag}_{p-q-q'}c_{p}d_{k},\\
[c^{\dag}_{p-q'}c_{p},c^{\dag}_{p'+q'}c^{\dag}_{p-q'}c_{p}c_{p'}]
=&[c^{\dag}_{p-q'}c_{p},c^{\dag}_{p-q'}c_{p}]c^{\dag}_{p'+q'}c_{p'}
  +c^{\dag}_{p-q'}c_{p}[c^{\dag}_{p-q'}c_{p},c^{\dag}_{p'+q'}c_{p'}]\\
=&c^{\dag}_{p-q'}c_{p}[c^{\dag}_{p-q'}c_{p},c^{\dag}_{p'+q'}c_{p'}]
=0,
\end{aligned}
\end{equation}
since in restricted Hilbert space (with cutoff in momentum space)
 we have
\begin{equation} 
\begin{aligned}
c^{\dag}_{p-q'}c_{p}d^{\dag}_{k+q}c^{\dag}_{p-q}c_{p}d_{k}|\psi\rangle
=&d^{\dag}_{k+q}c^{\dag}_{p-q-q'}c_{p}d_{k}|\psi\rangle,\\
d^{\dag}_{k+q}c^{\dag}_{p-q}c_{p}d_{k}c^{\dag}_{p-q'}c_{p}|\psi\rangle
=&0.
\end{aligned}
\end{equation}
In many-electron average, the second term in last line of above expression
becomes
\begin{equation} 
\begin{aligned}
\sum_{k,p,q}[\langle d^{\dag}_{k+q}d_{k}\rangle \langle c^{\dag}_{p-q-q'}c_{p}\rangle
-\langle d^{\dag}_{k+q}c_{p}\rangle \langle c^{\dag}_{p-q-q'}d_{k}\rangle],
\end{aligned}
\end{equation}
at zero-temperature.
As mentioned above, the
density operator $\hat{n}=\sum_{p}\psi^{\dag}_{p-q'}\psi_{p}$
is momentum $q'$-dependent due to the polaron-polaron interaction.
Then we write the current as $J=evn=ev\sum_{p-q'}N_{F}(\varepsilon_{p-q'})$,
i.e., the constant carrier concentration within the expression of
polaron-polaron interaction-induced current can be replaced by the distribution functon
at scatterred state $p-q'$.
In semiclassical Boltzmann transport theory, the 
distribution function here can be further replaced by $N_{F}(r,t,p-q')$,
which is simply denoted as $N_{F}$ in the following.

In the collisionless limit (collision free),
the above continuity equation should be rewritten as
\begin{equation} 
\begin{aligned}
\frac{\partial N_{F}}{\partial t}\bigg|_{coll}
=&
\frac{\partial N_{F}}{\partial t}
+{\bf v}\cdot\frac{\partial N_{F}}{\partial {\bf r}}
+{\bf F}\cdot\frac{\partial N_{F}}{\partial {\bf q}'}\\
=&
\frac{\partial N_{F}}{\partial t}
+{\bf v}\cdot\frac{\partial N_{F}}{\partial {\bf r}}
-\dot{{\bf q}'}\cdot\frac{\partial N_{F}}{\partial {\bf q}}=0,
\end{aligned}
\end{equation}
We consider here the transferred momentum during polaron-polaron interaction $q'$
is induced by the driving field with the force ${\bf F}=-\dot{{\bf q}}'=e{\bf E}$.
Then we have $\frac{\partial N_{F}}{\partial t}\bigg|_{coll}=
{\bf F}\cdot\frac{\partial N_{F}}{\partial q'}$
in a homogeneous (${\bf r}$ independent) and steady (time independent) system.
Note that in the presence of the driving field,
$\frac{\partial N_{F}}{\partial t}\bigg|_{coll}=0$ is correct only in the instantaneous 
and single collision approximation where the 
$\frac{\partial N_{F}}{\partial {\bf q}'}=0$
and Boltzmann transport equation becomes independent of the external force.
In nonequilibrium case,
the distribution function contains an additional term except the equilibrium distribution function,
$e\tau {\bf v}\cdot {\bf E}(\frac{-\partial N_{F}^{0}}{\partial \varepsilon_{k}})$,
in relaxation time approximation,
and both the collisions and dissipation drive the system back to equilibrium.

The changing rate of the distribution function can be rewritten as
\begin{equation} 
\begin{aligned}
\frac{\partial N_{F}}{\partial t}\bigg|_{coll}=
\sum_{p}[w_{p,p-q'}N_{F}(p)(1-N_{F}(p-q'))
-
w_{p-q',p}N_{F}(p-q')(1-N_{F}(p))],
\end{aligned}
\end{equation}
where $w_{a,b}$ denote the transition rate from state $a$ to state $b$
and this expression describes the
probability of an impurity scattered to the
polaronic state ($p-q'$) per unit time. 
For elastic scattering with scattering frequency $\omega=0$,
we have $w_{p-q',p}=w_{p,p-q'}$,
thus
\begin{equation} 
\begin{aligned}
\frac{\partial N_{F}(p-q')}{\partial t}\bigg|_{coll}
=&
{\bf F}\cdot\frac{\partial N_{F}(p-q')}{\partial ({\bf p}-{\bf q}')}\\
=&{\bf F}\cdot{\bf v}_{p-q'}\frac{\partial N_{F}(p-q')}{\partial \varepsilon_{p-q'}}\\
=&\sum_{p}w_{p,p-q'}(N_{F}(p)-N_{F}(p-q'))\\
=&\sum_{p}w_{p,p-q'}(N_{F}^{0}(p)+{\bf F}\cdot{\bf v}_{p}\tau_{p}\frac{-\partial N_{F}(p)}{\partial \varepsilon_{p}}
-N_{F}^{0}(p-q')-{\bf F}\cdot{\bf v}_{p-q'}\tau_{p-q'}\frac{-\partial N_{F}(p-q')}{\partial \varepsilon_{p-q'}})\\
=&\sum_{p}w_{p,p-q'}(N_{F}^{0}(p)-N_{F}^{0}(p-q'))\\
& +\sum_{p}w_{p,p-q'}({\bf F}\cdot{\bf v}_{p}\tau_{p}\frac{-\partial N_{F}^{0}(p)}{\partial \varepsilon_{p}}
-{\bf F}\cdot{\bf v}_{p-q'}\tau_{p-q'}\frac{-\partial N_{F}^{0}(p-q')}{\partial \varepsilon_{p-q'}}).
\end{aligned}
\end{equation}
Note that $e<0$ thus $|{\bf F}|<0$.
Since the first term is the chnaging rate in equilibrium distribution function
$\frac{\partial N^{0}_{F}}{\partial t}\bigg|_{coll}=0$,
this part of changing rate indeed does not contributes to the current $J$.
Thus we have
\begin{equation} 
\begin{aligned}
1=
\sum_{p}w_{p,p-q'}[\tau_{p-q'}
-\frac{{\bf v}_{p}}{{\bf v}_{p-q'}}
 \frac{\tau_{p}}{\tau_{p-q'}}
 \frac{\frac{\partial N_{F}^{0}(p)}{\partial \varepsilon_{p}}}{\frac{\partial N_{F}^{0}(p-q')}{\partial \varepsilon_{p-q'}}}
].
\end{aligned}
\end{equation}
For inelastic scatering with nonzero scattering frequency,
the detailed balance relation reads
 $w_{p-q',p}N_{F}^{0}(p-q')(1-N_{F}^{0}(p))=w_{p,p-q'}N_{F}^{0}(p)(1-N_{F}^{0}(p-q'))$,
similarly we can obtain
\begin{equation} 
\begin{aligned}
1=
\sum_{p}w_{p-q',p}\frac{1-N_{F}^{0}(p)}{1-N_{F}^{0}(p-q')}
[\tau_{p-q'}-\tau_{p}
{\rm cos}\theta_{p-q',p}
             \frac{N_{F}^{0}(p-q')}{N_{F}^{0}(p)}
             \frac{1-N_{F}^{0}(p-q')}{1-N_{F}^{0}(p)}
],
\end{aligned}
\end{equation}
where the relation 
$\frac{-\partial N_{F}^{0}(p-q')}{\partial \varepsilon_{p-q'}}
=N_{F}^{0}(p-q')(1-N_{F}^{0}(p-q'))$ is used.
$\theta_{p-q',p}$ is the angle between $p-q'$ and $p$,
and ${\rm cos}\theta_{p-q',p}=\frac{{\bf v}_{p-q'}\cdot{\bf v}_{p}}{|{\bf v}_{p-q'}||{\bf v}_{p}|}
=\frac{{\rm cos}\theta_{\hat{{\bf F}},p}}{{\rm cos}\theta_{\hat{{\bf F}},p-q'}}$,
with $\hat{{\bf F}}={\bf F}/|{\bf F}|$ is the direction projection of force.
In conclusion,
for both elastic and inelastic scattering,
we have $\frac{\partial N_{F}^{0}(p-q')}{\partial t}\bigg|_{coll}$
and $\frac{\partial N_{F}(p-q')}{\partial t}\bigg|_{coll}=
{\bf F}\cdot
\frac{\partial N_{F}(p-q')}{\partial (p-q')}\approx
{\bf F}\cdot
\frac{\partial N_{F}^{0}(p-q')}{\partial (p-q')}
={\bf F}\cdot {\bf v}_{p-q'}
\frac{\partial N_{F}^{0}(p-q')}{\partial \varepsilon_{p-q'}}$.
Since the scattering (collision) and dissipation tends to drive the system toward steady state,
in collisionless regime ($\omega\gg k_{B}T$) we have
 $\frac{\partial N_{F}(p-q')}{\partial t}\bigg|_{coll}
=\frac{\partial N_{F}^{0}(p-q')}{\partial t}\bigg|_{coll}=0$,
thus we have $\frac{\partial N_{F}^{0}(p-q')}{\partial \varepsilon_{p-q'}}=0$
thus at lease for the intraband longitudinal optical conductivity
where 
$\frac{N_{F}(sp)-N_{F}(sp')}{\varepsilon_{sp}-\varepsilon_{sp'}}
\approx \frac{\partial N_{F}(sp)}{\partial \varepsilon_{sp}}$ ($q'\rightarrow 0$),
we have ${\rm Re}\sigma_{ii}=0$.
The term 
$\frac{\partial N_{F}(p-q')}{\partial \varepsilon_{p-q'}}$
is also contained in both the normal conductivity and ordinary Hall conductivity.
However, at the regime of $\Lambda\gg\omega\gg k_{B}T$, 
the optical conductivity could tends to a frequency and UV cutoff-independent 
universal interband value $\sigma_{0}=e^{2}/4\hbar$\cite{Abedinpour S H}.
In the presence of strong external field,
both the mobility and conductivity are field-dependent\cite{Huang D},
and then the approximated Hall current (perpendicular to the applied electric field as well as
its magnetization) ${\bf J}=\sigma{\bf E}+\sigma_{H}{\bf H}\times{\bf E}$
(e.g., $J_{y}(x)=\sigma_{xy}E_{x}=ev\sum\psi^{\dag}(x)\tau_{y}\psi(x)$
where $\tau_{y}$ is the Pauli matrix
and $\psi(x)=e^{-ip_{y}y/\hbar}\psi(x,y)$ is the eigenvector although $p$ is nomore a good quantum number
in the presence of quenched random impurities)
where $\sigma_{H}$ is the Hall conductivity (transverse),
or even the dissipationless equilibrium anomalous Hall current\cite{Silvestrov P G,Lee W L} can exist.

The normal current and Hall current read
\begin{equation}
\begin{aligned}
J=&\int\frac{d^{3}(p-q')}{(2\pi)^{3}}
eve{\bf E}\cdot{\bf v}\tau
\left(-\frac{\partial N_{F}(p-q')}{\partial \varepsilon_{p-q'}}\right),\\
J_{H}=&\int\frac{d^{3}(p-q')}{(2\pi)^{3}}
ev
[e{\bf E}\cdot{\bf v}\tau(-\frac{\partial N_{F}^{0}(p-q')}{\partial \varepsilon_{p-q'}})
 -e(\frac{e\tau}{mc}{\bf H}\times{\bf E}\cdot {\bf v}\tau
\left(-\frac{\partial N_{F}^{0}(p-q')}{\partial \varepsilon_{p-q'}}\right)
],
\end{aligned}
\end{equation}
and the consequent normal and Hall conductivity are
\begin{equation}
\begin{aligned}
\sigma/\sigma_{0}=&\int\frac{d^{3}(p-q')}{(2\pi)^{3}}
e^{2}v^{2}\tau
\left(-\frac{\partial N_{F}(p-q')}{\partial \varepsilon_{p-q'}}\right),\\
\sigma_{H}/\sigma_{0}=&\int\frac{d^{3}(p-q')}{(2\pi)^{3}}
e^{2}v^{2}
[-(\frac{e\tau|{\bf H}|}{mc}\tau
\left(-\frac{\partial N_{F}^{0}(p-q')}{\partial \varepsilon_{p-q'}}\right)
].
\end{aligned}
\end{equation}
In zero-temperature limit,
$\left(-\frac{\partial N_{F}(p-q')}{\partial \varepsilon_{p-q'}}\right)
\rightarrow
\delta(\varepsilon_{p-q'}-\varepsilon_{F})$,
thus the conductivities become
\begin{equation}
\begin{aligned}
\sigma/\sigma_{0}=&
e^{2}v^{2}\tau
\rho(\varepsilon_{F}),\\
\sigma_{H}/\sigma_{0}=&
e^{2}v^{2}
[-(\frac{e\tau|{\bf H}|}{mc}\tau
\rho(\varepsilon_{F})
],
\end{aligned}
\end{equation}
where $\rho(\varepsilon_{F})=\int\frac{d^{3}(p-q')}{(2\pi)^{3}}
\delta(\varepsilon_{p-q'}-\varepsilon_{F})
=\int\frac{d^{3}(p)}{(2\pi)^{3}}
\delta(\varepsilon_{p}-\varepsilon_{F})$
is the density-of-states at Fermi energy.
The corresponding mobilities are can be obtained as $u=\sigma/ne$\cite{Barker Jr R E}
and at finite temperature the carrier concentration should
be replaced by the Boltzmann momentum distribution density operator
in terms of the partition function in the denominator 
$\frac{e^{-H_{int}/T}}{\sum_{p-q'}e^{-H_{int}/T}}c^{\dag}_{p-q'}c_{p}$,
or $\frac{e^{-H_{int}/T}}{{\rm Tr}e^{-H_{int}/T}}c^{\dag}_{p-q'}c_{p}$
in the orthonormal basis set.

As can be seen from the spectral functions
for $\alpha>1$ obviously deviate the delta function even at zero temperature (Fig.5)
which is due to the finite imaginary part of self-enegy (continuum region).
While for $\alpha=1$,
the quasiparticles are with sharp momentum in the presence of small disorder parameter.
In many-polaron system, only
when the random quanched impurities are not taken into account,
the effect of inpurity-majority interaction could be elastic especially at low temperature,
in which case the system acts like a dissipationless stationary state
and the Fermi-liquid description is valid.
This is different from the inelastic scattering events like the
electron-phonon coupling (with strong momentum dissipation on lattice) or the Raman scattering.
In the spectral function or RPA dynamical structure factor,
when the incoherence (continuum) part has large width,
the non-Fermi-liquid feature is obvious
and the momentum distribution function becomes continuous.

\subsection{WF law}

As most of the situations we discussed above is the Fermi-liquid case even
in the presence of weakly broadened momentum relaxation (${\rm Im}\Sigma(p)$;
see the spectral function in Fig.7),
the Wiedemann-Franz (WF) law should be obeyed in low-temperature limit,
where the Drude (intraband) conductivity in the presence of long relaxation,
in terms of static current-current propagator $\Pi_{ij}$, reads
\begin{equation}
\begin{aligned}
\sigma_{ij}(\omega)=\frac{\Pi_{ij}}{-i\omega+\frac{1}{\tau}}
=\frac{\Pi_{ij}\tau}{-i\omega\tau+1}=\frac{\sigma_{dc}}{-i\omega\tau+1},
\end{aligned}
\end{equation}
where the transport relaxation time has $\tau={\rm Im}\Sigma$ and 
$\sigma_{dc}\sim \Pi_{ij}\tau$ is the static Drude conductivity,
and
the temperature-dependence of thermal (heat) conductivity then reads
\begin{equation}
\begin{aligned}
\kappa\sim
\frac{n}{T}\int v^{2}\varepsilon^{2}\frac{\Pi_{ij}}{{\rm Im}\sigma}
\sim nT\Sigma_{ij}(\omega\rightarrow 0).
\end{aligned}
\end{equation}
Note that the elastic (randomly) quenched impurities will not leads to the breakdown of WF law,
although the system exhibits non-Fermi-liquid feature when we choose momentum eigenstate representation
due to the breakdown of momentum conservation
(while the energy conservation remains).
In the absence of neutral degrees of freedom and inelastic scattering,
the electrical and thermal (heat) relaxation rates are equal,
and the WF law is valid in both the $T\ll T_{F}$ and $T\gg T_{F}$ regions.
Here $T_{F}$ is the Fermi-temperature, which increases with the increasing impurity potential
and carrier density $n$,
and thus we have $T_{F}\sim\mu_{i}$ here.
In the absence of impurity,
$T_{F}=0$.
In these two temperature regions the conductivities obtained in small-$q$ limit
are $\sigma\sim T^{0}(T\ll 0.2)$ and $\sigma\sim T^{-1/2}(T\gg 0.2)$, respectively, as shown in Fig.17,
and then the corresponding thermal conductivities can be obtained\cite{Lavasani A}.
While for the case of larger scattering strength $g_{b}$ and higher polaron density,
the low-temperature resistivity should increased by $g_{b}$ in the $T<T_{F}$ region
in the absence of phonon scattering.

In the presence of
additional neutral degrees of freedom, such as thermal conductivity,
and the inelastic scattering due to the coupling with neutral excitations, 
like the nonequilibrium phonons or neutral collective modes,
the phonon scattering effects (including the linear high-temperature resistivity)
emerges only at temperature much higher than the Bloch-Gruneisen (or Debye) temperature,
which is $T_{BG}(T_{D})\sim n_{i}^{1/3}$
with $n_{i}$ the density of impurities which coupled with the phonons.
In high-Fermi temperature systems, like the metals,
when the temperature is much higher than the Bloch-Gruneisen (or Debye) temperature
(assumed higher than the superconductivity critical temperature here),
but lower than the Fermi-temperature,
the linear-in-temperature electrical resistivity emerges induced by the quasielastic phonon-scattering
(with phonon energy $\Omega\ll T_{F}\sim \mu_{i}$),
in which case the WF law is thus well obeyed
as long as the energy transfer does not comparable with the temperature scale.
Note that here quasielastic scattering again corresponds to the energy conservation but not the 
momentum conservation,
and for systems like the incoherent heavy fermion metals\cite{Chowdhury D,Varma C M} with strong interaction 
which exhibit linear-in-$T$ behavior,
the momentum would be strongly dissipated (with the relaxing current).
The linear-in-$T$ resistivity can be viewed as a reliable signature of the emergence of non-Fermi-liquid,
no matter the WF is obeyed or violated.
The existence of bosonic scattering mechanism (like the phonon or paramagnon scattering)
will leads to failure of WF law and non-Fermi liquid.
While in the opposite case with temperature much lower than the Fermi-temperature,
the WF law and Fermi-liquid theory are valid.
And in this case the Fermi-surface is larger when the polaronic coupling is weak enough.
Since we consider the weakly disordered (due to the randomly quenched impurities) and low-carrier density case,
the Fermi-temperature is low and the linear-in-$T$ resistivity is unrealizable in $T\ll T_{F}$ region
(with negligible phonon effect).

In Fermi-liquid system,
the valid WF law with a well-defined Fermi surface can be found 
only at the temperature much lower than the Fermi temperature,
in which case the quenched impurities' elastic scattering (with energy conservation but not momentum conservation)
is dominating (play a main role during both the charge and heat transports).
And in this case the WF law is valid as long as the polaron (as an electronic quasiparticle
due to the low-energy scattering) is existing and long-lived (with small ${\rm Im}\Sigma$).
However, due to the nonzero ${\rm Im}\Sigma$ and momentum transfer,
the total momentum is not conserved unlike some non-Fermi-liquid case.
While for the case with linear-in-$T$ resistivity and linear-in-$T$ scattering rate
when the temperature is much higher than the Bloch-Gruneisen (or Debye) temperature
in the so-called energy equipartition regime,
although the momentum transfer could still very large,
the energy transfer would always much lower than the temperature scale,
and it is thus treated quasielastic.
This case is recognized as non-Fermi-liquid with WF law valid.
The linear-in-$T$ behavior could also be found at a lower temperature scale in the
strongly interacting case, like in heavy fermion metal
with the incoherence (broadened) spectral peak,
but this is recognized as non-Fermi-liquid with WF law invalid.

As evidenced by the spectral functions presented above,
with the increase of dispersion parameter $\alpha$,
the non-Fermi-liquid feature, like the finite lifetime of polaron (as a long-lived quasiparticle),
will emerges guadually.
In this premise,
the WF law will be valid even in the presence of phonon modes
since the charge and heat transport are dominated by the long-lived polarons.

Note that although the long-lived polarons discussed here
has a rather small imaginary part of self-energy,
it is formed by the mobile impurities here,
and thus has an energy transfer larger than that from a randomly distributed impurities system
where the impurities are fixed,
which leads to a larger inverse quasiparticle lifetime compared to the inverse disorder lifetime,
${\rm Im}\Sigma>\frac{1}{\tau_{diso}}=2\pi n_{i}U^{2}\rho(\omega)$ where $U$ denotes the disorder potentials. 

\subsection{Momentum distribution function}

Except the damping rate (imaginary part of self-energy),
the feature of Fermi-liquid or non-Fermi-liquid can be seen from the absence of discontinuity 
of momentum distribution function (corresponds to an well- or ill-defined Fermi surfaces),
which reads
\begin{equation}
\begin{aligned}
N(p)=\int^{\infty}_{-\infty}\frac{d\omega}{2\pi} N_{F}(p,\omega,T)A(p,\omega),
\end{aligned}
\end{equation}
whose corresponding operator $\hat{N}_{p}=\langle c^{\dag}_{p}c_{p}\rangle$ 
is normalized and thus the polaron concentration reads
\begin{equation}
\begin{aligned}
n=\int^{p_{F}}_{0}N(p)dp=\int^{\mu_{i}}_{0}\rho(\omega)d\omega
=\int^{p_{F}}_{0}dp\int^{\mu_{i}}_{0}d\omega A(p,\omega).
\end{aligned}
\end{equation}

Due to the Pauli exclusion,
the $N(p)\le 1$ and becomes a Heaviside step function (corresponds to equilibrium impurity distribution)
when the spectral function
is exactly a $\delta$-function
which contains all the weight of spectrum (at zero temperature).
Note that
$N(p\ll p_{F}={}^{3}\sqrt{6\pi^{2}n})\approx 1$ requires the range of integration over the frequency
from positive infinite to negative infinite.
To calculate the momentum distribution function,
we make the approximation to self-energy $\Sigma_{0}(p,\omega)$ within the spectral function as,
\begin{equation}
\begin{aligned}
{\rm Re}\Sigma_{0}(p,\omega)\approx&
1/B^{3}ng_{b} \pi (-2 B^{2} q_{low}^{2} + 4 B q_{low} \mu_{m} - 4 B^{2} q_{low} p \\
&- 
   4 B \mu_{m} q_{up} + 4 B^{2} p q_{up} + 2 B^{2} q_{up}^{2} + 4 B q_{low} \omega - 4 B q_{up} \omega)+O(\omega^{2}),\\
{\rm Im}\Sigma_{0}(p,\omega)\approx&
1/B^{32} ng_{b} \pi (2 B q_{low} z - 2 B q_{up} z)+O(\omega^{2}).
\end{aligned}
\end{equation}
where we use the replacement $i\omega\rightarrow
\omega+iz$ and $q_{up}(q_{low})$ denotes the upper (lower) limit of transferred momentum during the
integral.
This implies that for a small value of nearly constant transport scattering rate,
we have $1/\tau_{tr}=2{\rm Im}\Sigma_{0}\propto 2ng_{b} \pi$.
That is also in consistent with the Eq.(10).

For large ${\rm Im}\Sigma(p,\omega)$,
the spectral function approaches a Lorentzian with half-width equals to ${\rm Im}\Sigma(p,\omega)=n\sigma_{c}$
where $\sigma_{c}$ is the scattering cross section.
However, even when ${\rm Im}\Sigma(p,\omega)=0$,
the momentum distribution function deviates from the Heaviside step function
as long as $T\neq 0$,
which can be seen from Fig.18,
as obtained by the following expression of momentum distribution function in the case ${\rm Im}\Sigma(p,\omega)=0$
\begin{equation}
\begin{aligned}
 N(p)
=&\int^{\infty}_{-\infty}
 d\omega N_{F}(\omega)\delta(\omega-(Ap^{\alpha}-\mu_{i}+ng_{b}))\\
=&\int^{\infty}_{-\infty}
 d\omega \frac{1}{1+e^{\frac{\omega}{T}}}
 \delta(\omega-(Ap^{\alpha}-\mu_{i}+ng_{b}))\\
\approx &Z(p) N_{F}(\varepsilon_{p}-\mu_{i}),
\end{aligned}
\end{equation}
where the last line can be obtained by using the coherence part of spectral function
presented below.
To prevent the issue pointed out by ref.\cite{Takada Y},
we use the wave function in first-order variational Chevy Ansatz,
which for single impurity and majority particle reads
\begin{equation}
\begin{aligned}
|\psi\rangle_{p}=(\phi_{0}c_{p}^{\dag}+\sum_{q}\phi_{q}c^{\dag}_{p-q}d^{\dag}_{q})|0\rangle_{\uparrow},
\end{aligned}
\end{equation}
where the parameters are
\begin{equation}
\begin{aligned}
\phi_{0}=&\frac{g_{b}\chi}{E-\varepsilon_{p}}=\frac{g_{b}\chi}{\Sigma_{p}}\\
        =&\frac{1}{\sqrt{1+(\frac{\Sigma_{p}}{E-(\varepsilon_{p-q}+\varepsilon_{q})})^{2}}},\\
\phi_{q}=&\frac{g_{b}\chi}{E-(\varepsilon_{p-q}+\varepsilon_{q})}\\
        =&\frac{\Sigma_{p}}{E-(\varepsilon_{p-q}+\varepsilon_{q})}\phi_{0}\\
        =&\frac{\Sigma_{p}}{E-(\varepsilon_{p-q}+\varepsilon_{q})}
\frac{1}{\sqrt{1+(\frac{\Sigma_{p}}{E-(\varepsilon_{p-q}+\varepsilon_{q})})^{2}}}.
\end{aligned}
\end{equation}
In the presence of minimized self-consistent polaron energy $E$,
the normalization condition 
$\langle\psi|\psi\rangle=|\phi_{0}|^{2}+\sum_{q}|\phi_{q}|^{2}=1$ is satisfied,
with another form of self-energy
$\Sigma_{p}=\frac{1}{\sqrt{\frac{1}{g_{b}^{2}\chi^{2}}-\frac{1}{E-(\varepsilon_{p-q}+\varepsilon_{q})}}}$.
While in the case of multi-impurity and multi-majority particle
(with $N_{i}$ impurity and $N_{m}$ majority particles),
we have
\begin{equation}
\begin{aligned}
|\psi\rangle_{p}=(\sqrt{N_{i}}\phi_{0}c_{p}^{\dag}+\sqrt{N_{m}}\sum_{q}\phi_{q}c^{\dag}_{p-q}d^{\dag}_{q})|0\rangle_{\uparrow},
\end{aligned}
\end{equation}
where the parameters are
\begin{equation}
\begin{aligned}
\phi_{0}=&\frac{N_{m}g_{b}\chi}{E-\varepsilon_{p}}=\frac{N_{m}g_{b}\chi}{\Sigma_{p}}\\
        =&\frac{N_{m}}{\sqrt{1+(\frac{\Sigma_{p}}{E-(\varepsilon_{p-q}+\varepsilon_{q})})^{2}}},\\
\phi_{q}=&\frac{N_{i}g_{b}\chi}{E-(\varepsilon_{p-q}+\varepsilon_{q})}\\
        =&\frac{N_{i}\Sigma_{p}}{E-(\varepsilon_{p-q}+\varepsilon_{q})}\phi_{0}\\
        =&\frac{N_{i}\Sigma_{p}}{E-(\varepsilon_{p-q}+\varepsilon_{q})}
\frac{N_{m}}{\sqrt{1+(\frac{\Sigma_{p}}{E-(\varepsilon_{p-q}+\varepsilon_{q})})^{2}}},
\end{aligned}
\end{equation}
where we define $\chi=\phi_{0}+\sum_{q}\phi_{q}$,
and similarly we have $\langle\psi|\psi\rangle=
N_{i}|\phi_{0}|^{2}+N_{m}\sum_{q}|\phi_{q}|^{2}=1$
with 
$\Sigma_{p}=\frac{N_{i}}{\sqrt{\frac{1}{g_{b}^{2}\chi^{2}}-\frac{N_{m}}{E-(\varepsilon_{p-q}+\varepsilon_{q})}}}$.

With the help of these parameters,
we can then rewrite the spectral function as
\begin{equation}
\begin{aligned}
A(\omega)\sim Z_{FL}\delta(\omega-(\varepsilon_{p}+\Sigma(p,\omega)))
+\sum_{q}|\phi_{q}|^{2}\delta(\omega-(\varepsilon_{p-q}+\varepsilon_{q}+\Sigma(p,\omega))),
\end{aligned}
\end{equation}
where the first term is the coherence part with the $\delta$-peak shifted by polaron energy,
and the second term is the incoherence part.
$Z_{FL}=|\phi_{0}|^{2}$ is the quasiparticle weight (residue)
in Fermi-liquid theory.
Or for the multi-particle case
\begin{equation}
\begin{aligned}
A(\omega)\sim N_{i}Z_{FL}\delta(\omega-(\varepsilon_{p}+\Sigma(p,\omega)))
+N_{m}\sum_{q}|\phi_{q}|^{2}\delta(\omega-(\varepsilon_{p-q}+\varepsilon_{q}+\Sigma(p,\omega))),
\end{aligned}
\end{equation}
Note that the spectral functions here correspond to the transition rate from the impurity vacuum to 
the polaron state.

Thus when consider only the coherent part for many-polaron case, and to second-order self-energy,
we can approximately obtain 
\begin{equation}
\begin{aligned}
A_{p\rightarrow p'}(\omega)=2\pi N_{i}g_{b}^{2}\delta(\omega-(\varepsilon_{p}+\Sigma(p,\omega))),\\
A_{p'\rightarrow p}(\omega)=2\pi N_{i}g_{b}^{2}\delta(\omega+(\varepsilon_{p}+\Sigma(p,\omega))),
\end{aligned}
\end{equation}
and thus the balance condition reads
\begin{equation}
\begin{aligned}
A_{p\rightarrow p'}(\omega)=A_{p'\rightarrow p}(-\omega).
\end{aligned}
\end{equation}
This conclusion is based on the assumption that the coupling $g_{b}$ is independent of the momentum,
and thus $g_{b}^{p\rightarrow p'}=g_{b}^{p'\rightarrow p}$.
And this conclusion is also valid for single impurity case.
While when $g_{b}^{p\rightarrow p'}\neq g_{b}^{p'\rightarrow p}$,
we obtain the detailed balance condition in grand caconical ensemble as
\begin{equation}
\begin{aligned}
\frac{A_{p\rightarrow p'}(\omega)}{A_{p'\rightarrow p}(-\omega)}
=&\frac{e^{-E_{0}/T}}{Z_{E_{0}}}\frac{Z_{E}}{e^{-(\varepsilon_{p}+\Sigma+E_{0})/T}}\\
=&\frac{\sum e^{-(\varepsilon_{p}+\Sigma+E_{0})/T}}{e^{-(\varepsilon_{p}+\Sigma+E_{0})/T}}\\
\end{aligned}
\end{equation}
where $E=\varepsilon_{p}+\Sigma+E_{0}$
and $E_{0}=0$ is the ground state energy and $Z_{E}$ is the partition function,
i.e., the detailed balance condition equals the ratio of Boltzmann-type densities
in impurity vacuum and in polaron state.

From Fig.18,
at finite temperature,
the power-law singularity emerges near the $\pm p_{F}$.
For both the even and odd $\alpha$,
the momentum distribution function shows particle-hole symmetry due to the low
electron density,
even 
in the places far away from Fermi surface.
Also, we found that, for $\alpha>1$,
there exists a range of momentum ($-p_{F}\sim p_{F}$;
which is slightly larger than the first Brillouin zone for noninteracting case) where $N(p)$ becomes a constant,
and this constant decreases with increasing temperature.
This is similar to, but with different mechanism,
the energy-distribution function studied in Ref.\cite{Bertrand C}
which is due to the bias voltage.
The discontinuity of noninteracting Fermi surface only appears at zero-temperature limit,
and the position of Fermi surface is modified as the $\alpha$ changes.
We note that, it seems the momentum distribution function in Fig.18
shows time-reversal asymmetry for even $\alpha$,
but this is purely a mathematical result but not physical result
and only the $p>0$ part is meanfull to detect.

Near the Fermi surface (the discontinuity of $N(p)$ at $p=p_{F}$),
the perturbed coherence part of Green's function reads
\begin{equation}
\begin{aligned}
G^{coh}(p,\omega)
=&\frac{1-N(p)}{\omega+i\eta-\varepsilon_{p}-{\rm Re}\Sigma(p,\omega)}\\
=&\frac{
1-\int^{\infty}_{-\infty}\frac{d\omega}{2\pi i}
\frac{e^{i\omega_{n}0^{+}}}{\omega+i\eta-\varepsilon_{p}-{\rm Re}\Sigma(p,\omega)}
}
{\omega+i\eta-\varepsilon_{p}-{\rm Re}\Sigma(p,\omega)},
\end{aligned}
\end{equation}
where $\eta=1/\tau={\rm Im}\Sigma(p,\omega)$ is the inverse lifetime of polaron.
From this we obtain a simpler way to calculate the momentum distribution function,
which integrate the frequency along the real axis\cite{Takada Y,Fratini E,Schindlmayr AF}
\begin{equation}
\begin{aligned}
N(p)\equiv
&\frac{\langle \psi|c_{p}^{\dag}c_{p}|\psi\rangle}{\langle\psi|\psi\rangle}\\
=&\int^{\infty}_{-\infty}\frac{d\omega_{n}}{2\pi i}
\frac{e^{i\omega_{n}0^{+}}}{\omega_{n}+i0-(Ap^{\alpha}-\mu_{i}+\Sigma(p,\omega_{n}))}\\
=&\int^{\infty}_{-\infty}\frac{d\omega_{n}}{2\pi i}
\frac{e^{i\omega_{n}0^{+}}}{\omega+i0-(Ap^{\alpha}-\mu_{i}+\frac{1}{g_{b}^{-1}-\sum_{q}\frac{1}{\omega+i0-\varepsilon_{p-q}-\varepsilon_{q}}})}\\
=&T\sum_{n}
\frac{e^{i\omega_{n}0^{+}}}{\omega_{n}+i0-(Ap^{\alpha}-\mu_{i}+T\sum_{m}
                    \int\frac{d^{3}q}{(2\pi)^{3}}
      T(p-q,\omega_{n}-\Omega_{m}))},
\end{aligned}
\end{equation}
where $\Sigma(p,\omega_{n})$ is the second-order contribution to the self-energy
at finite temperature.
$\psi$ is the responding electron wave function in real space
(like the Jastrow wave function).
As shown in Fig.20(d),
the integral over frequency of dressed impurity propagator provides the momentum distribution function,
i.e., the contour integration over all the occupied states.
While in the grand canonical ensemble,
the momentum distribution function can also be given by the Boltzmann-type one
with the real frequency Green's function
\begin{equation}
\begin{aligned}
N(p)
=&\int^{\infty}_{-\infty}\frac{d\omega_{n}}{2\pi }
\frac{e^{-\omega_{n}/T}}{\omega-(Ap^{\alpha}-\mu_{i}+\frac{1}{g_{b}^{-1}-\sum_{q}\frac{1}{\omega+i0-\varepsilon_{p-q}-\varepsilon_{q}}})},
\end{aligned}
\end{equation}
which is valid for $e^{-\varepsilon/T}\ll 1(\varepsilon\gg T)$ case.

As we stated above,
the $p_{F}={}^{3}\sqrt{6\pi^{2}n}$ 
is changeable with $n$ and thus will varies with temperature as well as self-energy.
Base on the approximated result of
mean-field shift at finite temperature $\Sigma_{0}=ng_{b}T$
(as verified in Sec.), 
the momentum distribution functions are obtained and shown in Fig.19,
where we found 
in $p>p_{F}$ region, the momentum distribution functions roughly obey the rule 
$N(p>p_{F})\sim e^{-cp}$ where $c>0$ is a parameter 
which decreases with increasing temperature.
The range of discontinuity (size of Fermi surface)
is reduced by the increasing temperature.
Also, we found that the speed of decrease of $N(p)$
with increasing $p$ is reduced when the temperature increase,
i.e.,
$\frac{dN(p)}{dp}\bigg|_{p=p_{F}}\sim\frac{1}{T}$,
which is in agreement with the Fermi-liquid theory.
However, we note that this relation is only valid in $p=p_{F}$
and becomes false in $p\neq p_{F}$ as obviously shown in Fig.19.
The size of Fermi surface is reduced as the temperature increases.
Although no shown in Fig.19,
our calculations further reveals that the value of $p_{F}$ (i.e., position of discontinuity)
increases (in a low speed) with increasing $|g_{b}|$ and $T$.
We note, through calculation, we found that the above results are still valid
when we use the second-order self-energy instead of the mean-field one,
no matter how large the parameters $\alpha$ and $\beta$ are.
In weak-coupling regime,
the Fermi-momentum is possible to be weakly modified by the temperature or coupling strength,
however,
in strongly correlated systems with electron density reaches or higher than half-filling,
the fastest decrease of $N(p)$ will occurs far away from the Fermi surface even at low-temperature limit,
such case is studied in, e.g., Ref.\cite{Singh R R P}.
The most obvious difference of the strongly correlated systems with the one discussed in this paper
is that the self-energy becomes linear in $\omega$ in large-$\omega$ region but not
becomes a constant.

For the case of low-temperature limit,
the following expression is also valid
\begin{equation}
\begin{aligned}
N(p)
=&\int\frac{d\omega}{2\pi}
\frac{e^{i\omega_{n}0^{+}}}{\omega_{n}+i0-(Ap^{\alpha}-\mu_{i}
+\int\frac{d^{3}q}{(2\pi)^{3}}
     \frac{d\Omega}{2\pi}
      T(p-q,\omega-\Omega))}.
\end{aligned}
\end{equation}
In the presence of free-energy
(or self-consistently obtained polaron energy\cite{Doggen E V H,Fratini E}),
the momentum distribution at large momentum ($p\gg p_{F}$)
can also be obtained as
$N(p\rightarrow\infty)=\frac{dE(g_{b})}{p^{4}dg_{b}^{-1}}$.

\section{Relation with the vertex correction}

As shown in Fig.20,
although they are all within the ladder approximation,
the difference between $T$-matrix and vertex correction 
is that the propagator of majority component appears in $T$-matrix is a particle propagator while
that appears in ladder diagram with vertex correction is a hole propagator
(with different sign of momentum compares to the particle one).
Note that even when the pair propagator discussed in this paper is the Cooper-channel one,
the superconductivity is impossible even at zero temperature due to the weak and instantaneous pairing interaction,
and the Cooper pair is harder to formed when the non-Fermi-liquid emerges (with quantum criticality).

In terms of imaginary time,
the particle propagator and hole propagator 
can be written as
\begin{equation}
\begin{aligned}
G_{q}=&-\sum_{k}\langle\mathcal{T}\{d_{k+q}(\tau),d^{\dag}_{k}(0)\}\rangle
=-\sum_{k}\langle\mathcal{T}\{d^{\dag}_{-(k+q)}(\tau),d_{-k}(0)\}\rangle,\\
G_{-q}=&-\sum_{k}\langle\mathcal{T}\{d_{-(k+q)}(\tau),d^{\dag}_{-k}(0)\}\rangle
=-\sum_{k}\langle\mathcal{T}\{d^{\dag}_{k+q}(\tau),d_{k}(0)\}\rangle.\\
\end{aligned}
\end{equation}
$\mathcal{T}$ is the time ordering operator (in terms of $\Delta \tau$)
which can also be replaced by the Boltzmann-type density operator when
we take the thermal average over states.
For the case of single impurity (hole),
we also have
\begin{equation}
\begin{aligned}
G_{q}=&-\sum_{p}\langle\mathcal{T}c_{p-q}(\tau)c^{\dag}_{p}(0)\rangle
=-\sum_{p}\langle\mathcal{T}c^{\dag}_{-(p-q)}(\tau)c_{-p}(0)\rangle,\\
G_{-q}=&-\sum_{p}\langle\mathcal{T}c_{-(p-q)}(\tau)c^{\dag}_{-p}(0)\rangle
=-\sum_{p}\langle\mathcal{T}c^{\dag}_{p-q}(\tau)c_{p}(0)\rangle,
\end{aligned}
\end{equation}
The Green's function in frequency domain and the 
retarded Green's function can be obtained by the relation
$G_{\pm q}(i\omega_{n})=G_{\pm q}(i(2n+1)\pi T)=\int^{i/T}_{0}d\tau G_{\pm q}e^{-\omega_{n}\tau}
=\int^{\infty}_{0}dtG^{R}_{q}e^{i\omega_{n}\tau}$.
The brackets in the above equations denote the time average here,
while for the configurational average over grand canonical ensemble
(thermal average), it can be relaced by
$\langle\cdot\cdot\cdot\rangle={\rm Tr}[\hat{n}_{m}\cdot\cdot\cdot]$,
where $\hat{n}_{m}$ is the density operator of majority particles
and the trace over all Fock states.
This is in the similar way with the dynamics of expectation values.
Usually, the pair propagator appears in $T$-matrix approximation while
the particle-hole bubble-like propagator appears in the dynamical polarization (beyond the RPA obviously
due to the ladder expansion).

Next we briefly discuss the vertex correction to the density-density correlation function
which appears if the pair propagator within the denominator of $T$-matrix is replaced by the particle-hole 
propagator.
Since the pair (particle-hole but not particle-particle) propagator $\Pi'(p,\omega)$
has the bubble form $\Pi'(p,\omega)
=\frac{d^{3}q}{(2\pi)^{3}}\frac{d\Omega}{2\pi}\Pi'(p-q,\omega-\Omega)
=\frac{d^{3}q}{(2\pi)^{3}}\frac{d\Omega}{2\pi}G_{\downarrow}(p-q,\omega-\Omega)G_{\uparrow}(q,\Omega)$,
where $G_{\downarrow(\uparrow)}$ denotes the impurity (majority) Green's function.
The two Green's function here have two poles on the opposite sides 
of real axis in frequency space, which make
the vertex function nonzero:
\begin{equation}
\begin{aligned}
\Pi(p,\omega)=&
\frac{d^{3}q}{(2\pi)^{3}}\frac{d\Omega}{2\pi}
G_{\downarrow}(p-q,\omega-\Omega)\Gamma(p-q,q;\omega-\Omega,\Omega)G_{\uparrow}(q,\Omega),\\
\Gamma(p-q,q;\omega-\Omega,\Omega)
=&1+n|g_{q,q'}|^{2}
G_{\downarrow}(p-q',\omega-\Omega')\Gamma(p-q',q';\omega-\Omega',\Omega')G_{\uparrow}(q',\Omega').
\end{aligned}
\end{equation}
$g_{q,q'}$ is the interaction vertex  
which depends only on the transferred momentum $q(q')$
for isotropic interaction.
But note that there
are some cases where the interaction vertex depends on both transferred and intrinsic momenta\cite{Marchand D J J}.
The vertex is related to the self-energy by the Ward identity
as $\Gamma(p-q,q;\omega-\Omega,\Omega)=1+\frac{\partial \Sigma(p)}{\partial p}$,
and for vertex in current-current correlation function,
it becomes
$\Gamma(p-q,q;\omega-\Omega,\Omega)=1+\frac{1}{v}\frac{\partial \Sigma(p)}{\partial p}
=1+\frac{\partial \Sigma(p)}{\partial \varepsilon_{p}}$.
As shown in the red bubble of Fig.20(c),
the polarization related to the scattering of majority particle
can be simply obtained as the RPA result,
$\Pi'(q,\Omega)=
\frac{N_{F}(\varepsilon_{k})(1-N_{F}(\varepsilon_{k+q}))(\varepsilon_{k+q}-\varepsilon_{k})}
{\Omega^{2}+(\varepsilon_{k+q}-\varepsilon_{k})^{2}}$.

\section{Conclusion}

In this paper,
we focus on the electronic properties as well as the electron correlations 
in a polaron system with tunable dispersion parameter in Fermi-liquid picture.
Both the single/many-polaron and zero/finite temperature cases are discussed
and compared,
and since we consider the low-temperature regime
with the dominating low-energy elastic scattering and weak polaronic coupling $g_{b}$,
the momentum (or currents) relaxation is very slow and can be treated as perturbation,
which is very different with the  strongly correlated systems including the heavy fermion
metals where strong interactions.
A most obvious difference is that the non-Fermi-liquid
behaviors, including the Cooper pairing instability and the phonon-induced linear-in $T$ 
resistivity are absence in the system we investigated in this paper.
This also guarantees the validity of Fermi-liquid theory,
as be verified in the observables like the
self-energy, spectral function, optical conductivity (both the planar and 3D one), free energy,
and momentum distribution function,
in this paper.
Also, we found that the enhancement of bare coupling $g_{b}$ indeed contributes very little to
the broadening of the peak of spectral function (i.e., non-Fermi-liquid state).

During the numerical calculations,
we set a series of dispersion parameters to both the impurity and majority particles,
and the effect of dispersion parameter (to the Fermi-liquid/non-Fermi-liquid
state) are revealed.
We found that for the systems with second-order (in bare coupling $g_{b}$) self-energy
exhibit almost prefect Fermi-liquid feature for low dispersion parameter $\alpha$
(at least for $\alpha<4$),while for $\alpha\ge 4$,
the peaks of sepctral function are only weakly broadened,
i.e., exhibit very weak non-fermi-liquid feature (even weaker than the marginal Fermi liquid),
and the
spectral functions in energy representation
have a narrower peak than that in momentum representation,
which means the energy $\omega$ is a better quantum number (more close to the exact eigenstate)
than the momentum $p$.
Overall, the spectral function exhibits very weak incoherent Lorentzian feature for systems
with second-order self-energy compares to the one with first order self-energy.
And the well preserved Fermi-liquid picture can also be seen from other physical quantities
as discussed in this paper.
The Fermi-liquid picture can also be verified by the imaginary
part of second-order self-energies which vanishing as $\omega^{c}(T^{c}),c\gg 1$ when close to
Fermi surface, in constrast with the first-order self-energies.

Except the effect of finite low temperature,
dispersion parameters $\alpha$ and $\beta$,
and the order of $g_{b}$ in self-energy,
the effects of IR and UV cutoff are also be revealed in the calculation of free-energy as well as
the optical conductivities.
Here we note that the polaron self-energy calculated in
our system is in some extent similar to the leading-order self-energy in the 
usual solid state condensed systems,
where the polaron is absence and thus the $T$-matrix is replaced by the corresponding boson self-energy
term,
and the first order self-energy in this paper 
($\Sigma\sim ng_{b}$ for zero temperature and $\Sigma\sim ng_{b}T$ for finite low temperature)
corresponds to the one with one loop boson self-energy correction,
while the second order self-energy in this paper 
($\Sigma=\int g_{b}Gg_{b}G'=Eq.(2)$ for zero temperature
and $\Sigma=T\int GT$ for finite low temperature)
corresponds to the one with two loop boson self-energy correction,
which is usually caused by the fermi-boson Yukawa coupling or the disorder scattering.

Also, the disorder induced by static random quenched impurities (when consider the many-polaron effect)
is usefull in studying the validity of WF law in our system.
Although the impurities here are mobile but not fixed
(and thus the energy transfer is slightly larger than the case with fixed impurities),
we verified that the WF law is valid here as long as the energy transfer is much lower than the $\mu_{i}$
($\mu_{i}\gg T$ here),
and thus both the momentum and currents (heat and charge) relax slowly 
(with the low-energy processes which can be treated as perturbations) 
and equally, with the dominating elastic scattering.
We also mentioned another case in non-Fermi-liquid state where the WF law is still valid
in the aid of phonons
when $T\gg T_{BG}(T_{D})$,
as the energy transfer is much lower than the temperature
and thus the process of momentum transfer to phonons relax the momenta and currents in a same rate.

We also want to note that, for 
the optical conductivities calculated in this paper,
we considered a polaronic momentum transfer $q$ which is very small,
and it is induced by an external driving field
which with a vanishing photon momentum,
and thus the selection rule is $\omega=s\varepsilon_{p}-s'\varepsilon_{p'}$.
Since the $q$ is small but finite,
the optical conductivity here is not the same with the normal ones which has a zero momentum transfer
(i.e., the optical limit).
That is why the
 intraband optical conductivity is nonzero even at zero temperature for energy larger than chemical potential,
while in the absence of polaronic momentum transfer,
the optical conductivity should contains only the interband part at zero temperature.
For the momentum distribution function calculated in this paper,
a discontinuity can alwways be found at $p=p_{F}$
which corresponds to a well defined Fermi surface as well as the preserved fermi-liquid state.
However, we found a broken time-reversal symmetry (TRS) in momentum distrubution functions with even $\alpha$,
this in fact need not to worry since the $p<0$ part in momentum distrubution functions
have not much physical meaning here.

\end{large}
\renewcommand\refname{References}

\clearpage

\begin{figure}
\centering
\begin{subfigure}
  \centering
  \includegraphics[width=0.3\linewidth]{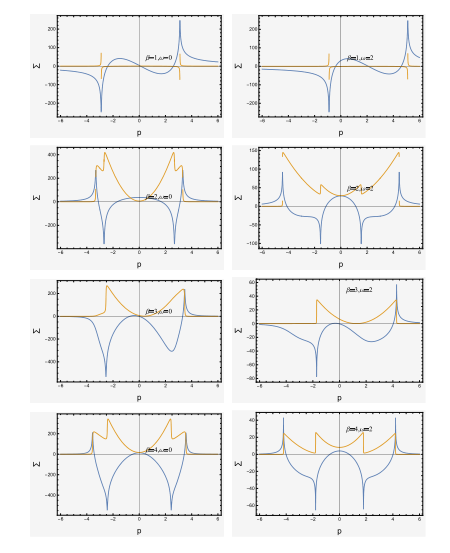}
\end{subfigure}
\caption{Self-energy calculated by Eq.(). The UV cutoff $\Lambda$ is set as 3 eV here and
the integral over momenta of majority particle is from $-\Lambda$ to $\Lambda$.
The model parameters are set as $A=1$ and $B=0.5$, with chemical potential $\mu_{i}=\mu_{m}=0.1$.
The blue and yellow lines represent the real part and imaginary part, respectively.
We integrate over the momentum of majority particle in a range of $-\Lambda$ to $\Lambda$
where the UV cutoff $\Lambda$ is set as 3.
Note that here the area factors ($(2\pi)^{d}$) are not taken into account in all panels.
}
\end{figure}

\clearpage
\begin{figure}
\centering
\begin{subfigure}
  \centering
  \includegraphics[width=0.3\linewidth]{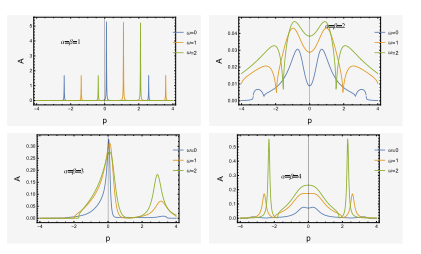}
\end{subfigure}
\caption{Spectral function base on the self-energy in Fig.1.
}
\end{figure}

\clearpage
\begin{figure}
\centering
\begin{subfigure}
  \centering
  \includegraphics[width=0.3\linewidth]{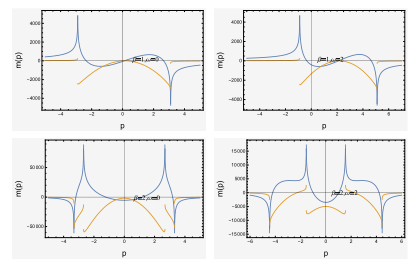}
\end{subfigure}
\caption{Effective mass in static approximation.
For simplicity, we set the mass of majority particle as a momentum-independent constant.
The blue and yellow lines represent the real part and imaginary part, respectively.
We integrate over the momentum of majority particle in a range of $-\Lambda$ to $\Lambda$
where the UV cutoff $\Lambda$ is set as 3.
}
\end{figure}

\clearpage
\begin{figure}
\centering
\begin{subfigure}
  \centering
  \includegraphics[width=0.2\linewidth]{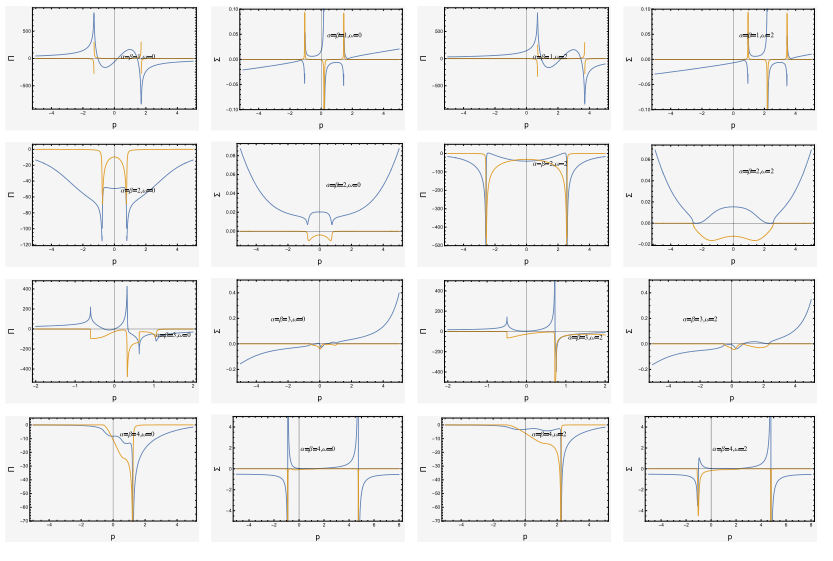}
\end{subfigure}
\caption{Self-energy calculated by Eq.().
}
\end{figure}

\clearpage

\begin{figure}
\centering
\begin{subfigure}
  \centering
  \includegraphics[width=0.3\linewidth]{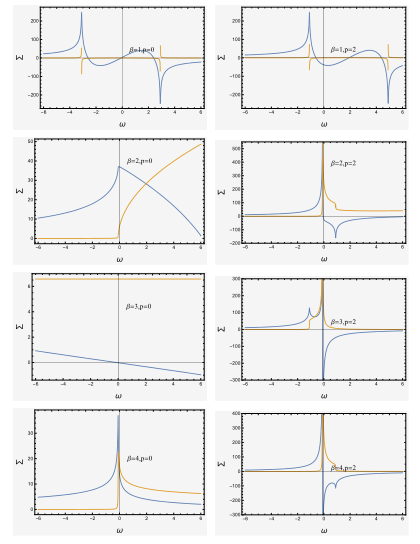}
\end{subfigure}
\caption{Pair-propagator and second-order contribution to self-energy 
as functions of $\omega$ for different $\alpha$ and $\beta$
and momenta $p$.
}
\end{figure}

\clearpage
\begin{figure}
\centering
\begin{subfigure}
  \centering
  \includegraphics[width=0.3\linewidth]{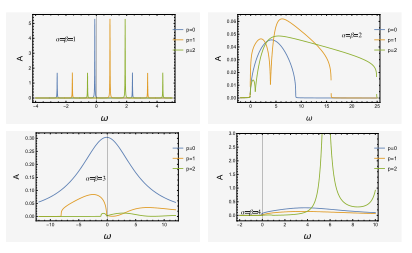}
\end{subfigure}
\caption{Spectral function calculated base on the first-order self-energy $\Sigma_{0}$ in Fig..
.
}
\end{figure}

\clearpage
\begin{figure}
\centering
\begin{subfigure}
  \centering
  \includegraphics[width=0.2\linewidth]{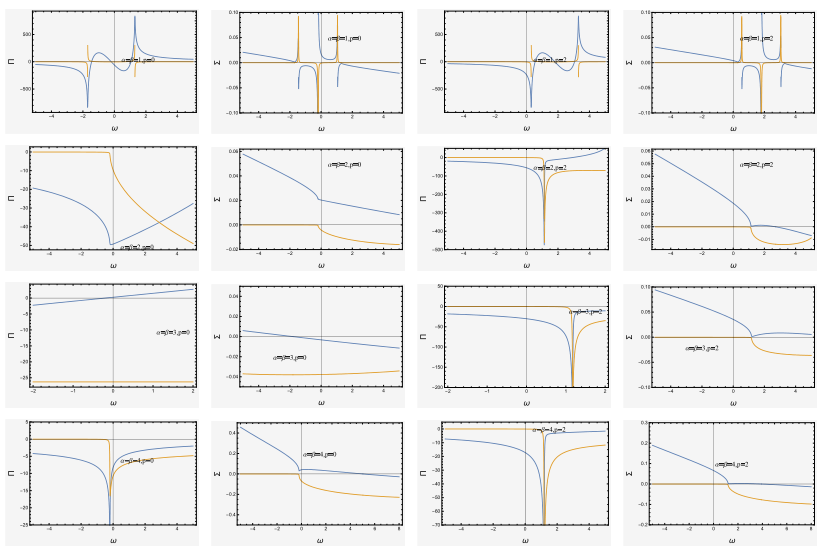}
\end{subfigure}
\caption{Pair propagator (a) and the interaction vertex ($T$-matrix) in strong interacting limit
with the UV cutoff very close to the chemical potential $\mu_{\downarrow}$.
}
\end{figure}

\clearpage
\begin{figure}
\centering
\begin{subfigure}
  \centering
  \includegraphics[width=0.3\linewidth]{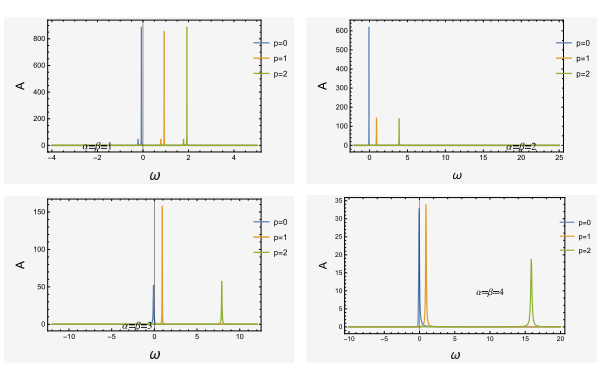}
\end{subfigure}
\caption{Spectral function calculated base on the second-order self-energy $\Sigma$ as a function of $\omega$.
}
\end{figure}

\clearpage
\begin{figure}
\centering
\begin{subfigure}
  \centering
  \includegraphics[width=0.3\linewidth]{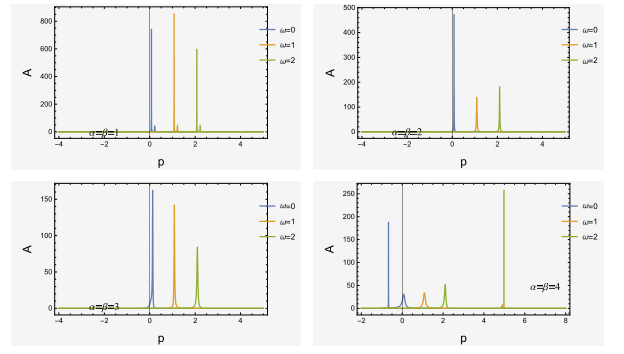}
\end{subfigure}
\caption{Spectral function calculated base on the second-order self-energy $\Sigma$ as a function of $p$.
}
\end{figure}

\clearpage
\begin{figure}
\centering
\begin{subfigure}
  \centering
  \includegraphics[width=0.3\linewidth]{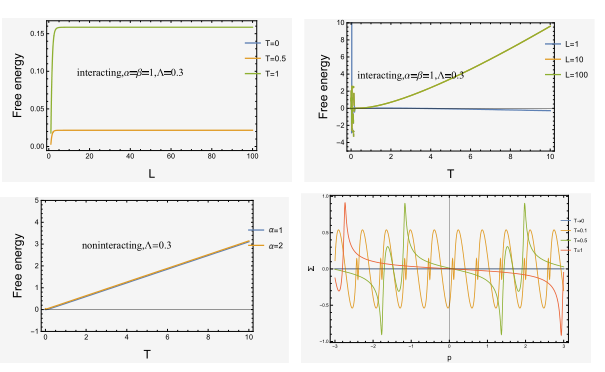}
\end{subfigure}
\caption{Free energy for interacting case and noninteracting case as a function of running length scale
$L$ and temperature $T$.
Note that during the calculation of free energy,
we neglect the Matsubara frequency of the impurity but focus only on its momenta.
The last panel shows the self-energy at finite temperature $T$ according to Eq.().
}
\end{figure}

\clearpage
\begin{figure}
\centering
\begin{subfigure}
  \centering
  \includegraphics[width=0.3\linewidth]{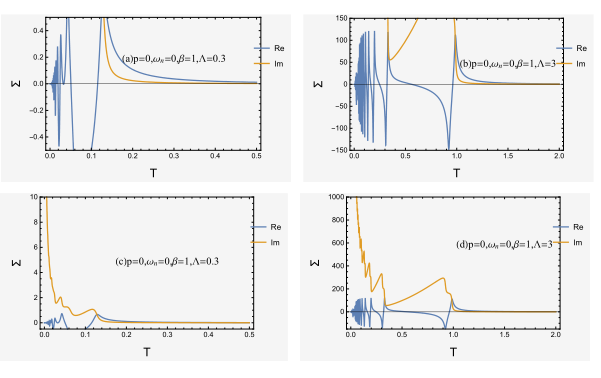}
\end{subfigure}
\caption{First-order contribution to self-energy $\Sigma_{0}$
as a function of temperature. Momentum cutoff is set as 0.3 and 3 in (a) and (b),
respectively.
(c) and (d) choose a different vertical axis scale compare to (a) and (b).
}
\end{figure}

\begin{figure}
\centering
\begin{subfigure}
  \centering
  \includegraphics[width=0.3\linewidth]{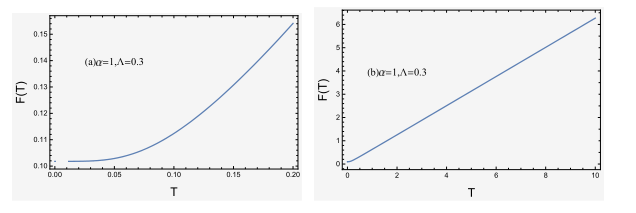}
\end{subfigure}
\caption{Free-energy calculated by applying the approximation of mean-field energy shift at finite temperature
$\Sigma_{0}=ng_{b}T$.
(a) and (b) correspond to a different choice of horizontal axis.
(b) reveals the linear behavior of free-energy at higher temperature
while (a) reveals the parabolic behavior.
}
\end{figure}

\clearpage
\begin{figure}
\centering
\begin{subfigure}
  \centering
  \includegraphics[width=0.3\linewidth]{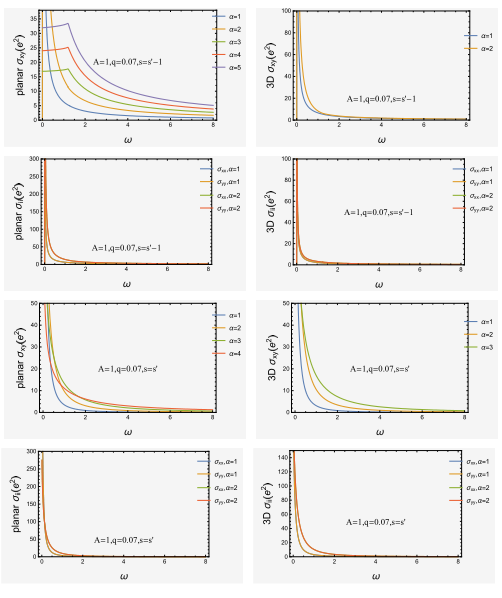}
\end{subfigure}
\caption{Planar and 3D transverse optical conductivities $\sigma_{xy}$ and $\sigma_{ii}(i=x,y)$
 with different dispersion order $\alpha$.
The four panels with $s=s'-1$ correspond to the interband transition while the 
four panels with $s=s'$ correspond to the intraband transition.
The transferred momentum is set in optical limit as $q=0.07=0.7\mu_{\downarrow}$.
We note that, except for the 3D interband optical conductivity (fourth panel),
we have $\sigma_{xx}=\sigma_{yy}$ with the same $\alpha$.
The range of integral of momentum $p$ is taken from zero to $\Lambda=3$.}
\end{figure}

\clearpage
\begin{figure}
\centering
\begin{subfigure}
  \centering
  \includegraphics[width=0.3\linewidth]{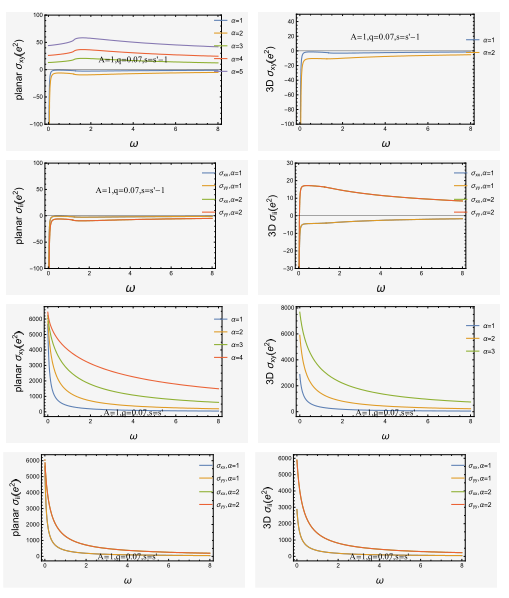}
\end{subfigure}
\caption{
Imaginary part of optical conductivities correpond to Fig..}
\end{figure}

\clearpage
\begin{figure}
\centering
\begin{subfigure}
  \centering
  \includegraphics[width=0.4\linewidth]{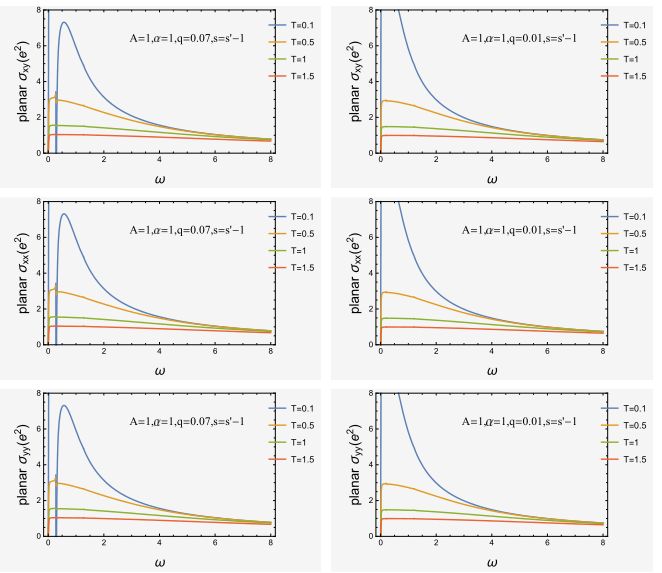}
\end{subfigure}
\caption{
Interband optical conductivity at finite temperature
with $\alpha=1$.
}
\end{figure}

\clearpage
\begin{figure}
\centering
\begin{subfigure}
  \centering
  \includegraphics[width=0.4\linewidth]{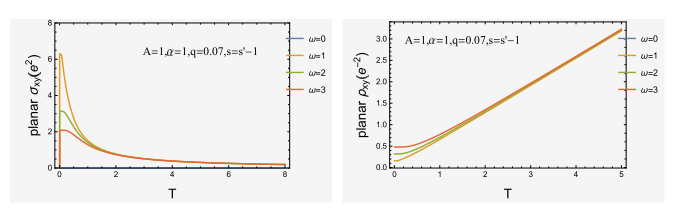}
\end{subfigure}
\caption{
Interband optical conductivity and the corresponding resistivity as a function of temperature.
}
\end{figure}

\clearpage
\begin{figure}
\centering
\begin{subfigure}
  \centering
  \includegraphics[width=0.4\linewidth]{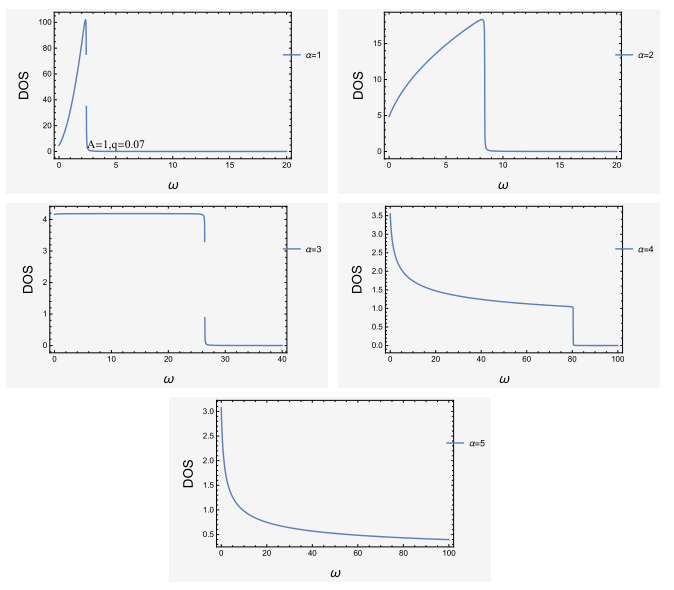}
\end{subfigure}
\caption{Density-of-states of polaron as a function of $\omega$
with different values of $\alpha$.At large enough $\omega$,
the density-of-states increases with the increasing $\alpha$.}
\end{figure}

\clearpage
\begin{figure}
\centering
\begin{subfigure}
  \centering
  \includegraphics[width=0.4\linewidth]{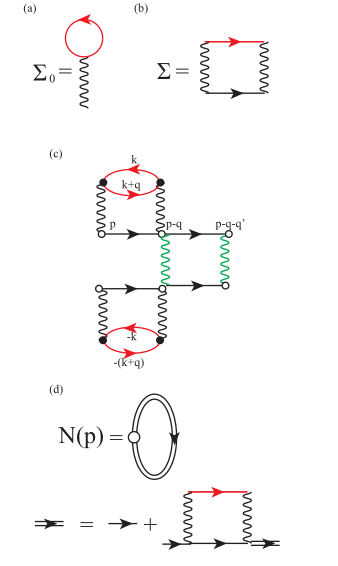}
\end{subfigure}
\caption{(a) The first-order contribution to polaron self-energy in the presence of vertex (Eq.()).
(b) The second-order contribution to polaron self-energy in the absence of vertices.
(c) The second-order contribution to polaron self-energy in the presence of vertices.
The red straight lines and black straight lines with arrowhead correspond to the propagators of majority 
particle and impurity, respectively.
The upper $T$-matrix contains the particle-hole scattering channel,
and the lower $T$-matrix contains the cooper-like scattering channel.
Many-polaron case is considered in (c),
where the black nodes denote the vertices which including the vertex correction.
while the white nodes denote the impurities where the external impurity
propagators are attached to.
The black and green wavy lines correspond to the $g_{b}$ and $g_{p-p}$, respectively.
(d) Momentum distribution function $N(p)$.
The while node here stands the term $c_{p}^{\dag}c_{p}$
and the double line stands the dressed Green's function.
Only the short-range interaction $g_{b}$ is considered here,
while the ring contribution as well as the exchange contribution\cite{Takada Y,Takada Y2} are not considered here.
}
\end{figure}

\clearpage
\begin{figure}
\centering
\begin{subfigure}
  \centering
  \includegraphics[width=0.4\linewidth]{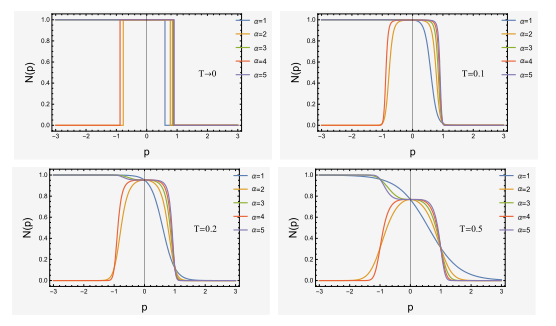}
\end{subfigure}
\caption{Momentum distribution function $N(p)$ at finite temperature
calculated by the $\delta$-type spectral function (Eq.()).}
\end{figure}

\clearpage
\begin{figure}
\centering
\begin{subfigure}
  \centering
  \includegraphics[width=0.4\linewidth]{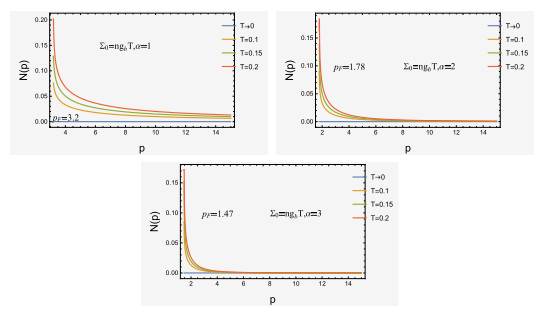}
\end{subfigure}
\caption{Momentum distribution function $N(p)$ at finite temperature
calculated by Eq.() in $p\ge p_{F}$ region.
The approximation of mean-field polaron energy at finite temperature $\Sigma_{0}=ng_{b}T$ is used here.}
\end{figure}

\end{document}